\input epsf
\input mn
\onecolumn


\def\bahat{\hat{\bf a}}
\def\bAhat{\hat{\bf A}}
\def\bbhat{\hat{\bf b}}
\def\bchat{\hat{\bf c}}
\def\bBhat{\hat{\bf B}}

\def\be{{\bf e}}
\def\br{{\bf r}}
\def\bR{{\bf R}}
\def\bF{{\bf F}}
\def\dv{\delta\dot\br}
\def\grad{\nabla}
\def\de{\delta e}
\def\rp{r_{\rm p}}

\def\etal{{\sl et al.}\ }
\def\bP{{\bf P}}


\begintopmatter

\title{The Effect of Encounters on the Eccentricity of Binaries in Clusters}

\author{Douglas C.~Heggie$^1$ and Frederic A.~Rasio$^2$}
\medskip
\affiliation{$^1$ Department of Mathematics and Statistics, University of Edinburgh, 
King's Buildings, Edinburgh EH9 3JZ, UK. Email: d.c.heggie@ed.ac.uk}
\smallskip
\affiliation{$^2$ Department of Physics, M.I.T., Cambridge, Massachusetts 02139, USA. 
Email: rasio@mit.edu}

\shortauthor{D.~C.~Heggie and F.~A.~Rasio}
\shorttitle{The Eccentricity of Binaries in Clusters}

\abstract{
We derive analytical expressions for the change in the orbital eccentricity of
a binary following a distant encounter with a third star on a hyperbolic or 
parabolic orbit. To establish the accuracy of these expressions,
we present detailed comparisons with the results of direct numerical integrations
of the equations of motion for the three bodies.
We treat with particular care the difficult case of a binary with 
zero initial eccentricity. In this case, we show that
the eccentricity $\delta e$ induced
by the encounter declines in general as a power-law, $\delta e\propto (a/\rp)^{5/2}$, where
$a$ is the binary semi-major axis and $\rp$ is the periastron distance of the encounter.
This power-law arises from the octupole-level secular perturbation of the binary. 
In contrast, non-secular quadrupole-level perturbations induce an eccentricity change
that declines exponentially with $\rp$. These non-secular effects
can become dominant at sufficiently small $\rp$, for a sufficiently high
relative velocity, or for a sufficiently massive perturber. 
We also derive cross sections for eccentricity change and
compare our results with those of previous studies based on numerical scattering
experiments. Our results have important implications for a number of 
astrophysical problems including, in particular,
the evolution of binary millisecond pulsars in globular clusters.
} 


\maketitle

\section{Introduction}

Three-body scattering plays an essential role in stellar
dynamics. In particular the interactions between binary stars and single
stars in globular clusters are thought to be an important dynamical
process, providing support against core collapse
(Spitzer 1987; Hut et al.\ 1992). What is important here is the energy
exchanged, but there are applications in which it is also important to
understand the changes in the eccentricity of a binary caused by an
encounter with another star. In some cases even very small changes in
eccentricity can be very significant. Most importantly, binary millisecond pulsars
have been discovered in large numbers in globular clusters  
and their orbital eccentricities can be measured with extraordinary
precision.
Although the orbits of binary millisecond pulsars in the galactic plane
are nearly circular 
(for example, PSR J2317+1439 has a measured eccentricity of $1.2\times10^{-6}$;
see Camilo \etal 1993), those found in the dense
stellar environment of globular clusters may be perturbed by the 
high rate of encounters with other stars. If, therefore, the
rate at which the eccentricity is changed by encounters is known,
interesting constraints can sometimes be placed on the lifetimes of the
binaries (Rappaport \etal 1989; Rasio \& Heggie 1995). 

Much of our knowledge of the effect of encounters on eccentricity is
numerical (Hills 1975; Hills \& Fullerton 1980; Hut \& Paczy\'nski 1984;
Rappaport \etal 1989). In the present paper the approach is
analytical, with numerical work used only to confirm the theoretical
results. Heggie (1975) gave a theoretical estimate (based on
first-order perturbation theory) for the change in eccentricity, and
found that it varies inversely as the $3/2$ power of $\rp$, the
distance of closest approach of the third body.  Unfortunately, his
result vanishes in the important case in which the initial orbit is
circular.  Phinney (1992) and Heggie (unpublished) then found, again
analytically, that in this case the induced eccentricity appeared to
decrease exponentially with a power of $\rp$, though the results were restricted
to special situations (e.g., equal masses, coplanar motions, and parabolic or
rectilinear motion of the third body).
While the authors of the present paper were reworking the theory for
more general cases, S.L.W.\ McMillan pointed out to us, on the basis of
numerical experiments with a new scattering package (McMillan 1996),
that the predicted exponential dependence for initially circular binaries
fails when the masses of the binary
components are unequal (which clearly is the important
case in applications).  This discovery stimulated much of the
remaining work for the present paper.

In this paper our aim is to clarify the regimes in which the various
theoretical results are applicable, and to extend them, where necessary,
to arbitrary masses and arbitrary geometries.  The one common assumption
is that the encounter is tidal, i.e., that $\rp\gg a$, where $a$ is the
initial semi-major axis of the binary, though it may be necessary to
take account of the masses of the participants in this condition when
the masses are very different.  We also usually assume that the
encounter is slow.

In the following section we outline the theoretical results, but leave
detailed derivations to an appendix.  We also compare the theoretical
results with new numerical data, designed to explore the full dependence
on $\rp$, the masses, the initial eccentricity of the binary, and the
geometry of the orbits.  In Section~3 we convert our results into cross
sections, which are then compared with results of other authors.   In
the last section we summarise our conclusions and outline briefly
some applications. Another summary, which places particular emphasis on
applications to binary pulsars in globular clusters, 
can be found in Rasio \& Heggie (1995).

\section{General Expressions for the Change in Eccentricity}

\subsection{Notations, Assumptions and Basic Equations}

After setting out our notation and assumptions, 
in this section we shall give the
basic equations for the perturbation of a binary by a third body.
Much of this is elementary, but it is helpful because it 
makes some properties of our detailed results quite obvious.

We consider an encounter between a single star of mass $m_3$ and a
binary whose components have masses $m_1$ and $m_2$, and we define
$M_{12}=m_1+m_2$ and $M_{123}=M_{12}+m_3$.  Long before the
encounter begins, the eccentricity of the binary is $e$, (though often
we shall assume this is zero), and its semi-major axis is $a$.  The
orbit of the third body relative to the barycentre of the binary is
hyperbolic with eccentricity $e^\prime$.  We denote its impact
parameter by $p$ and its speed relative to the barycentre, while they
are still far apart, by $V$. Then we have
$$
e^{\prime} = \sqrt{1 + {p^2V^4\over G^2M_{123}^2}},\eqno\stepeq
$$ 
and we shall be able to obtain results for a
parabolic encounter by letting $V\to0$, or $e^\prime\to1$.  

We make two assumptions about the encounter, that it is both {\sl tidal\/}
and {\sl slow\/}.  For the first of these, we assume that the closest
distance of approach of the third body, which will be denoted by $\rp$,
is still considerably larger than $a$.  The ratio of these distances
will play the role of a perturbation parameter.  For the second
assumption, we require that the angular velocity of the perturber, when
it is closest to the binary, is considerably smaller than the angular
velocity of the binary components.  It is easily shown that this
is equivalent to the condition
$$
\sqrt{{M_{12}\over M_{123}}} \left({\rp\over
a}\right)^{3/2}{1\over\sqrt{e^\prime+1}} \gg 1,\eqno\stepeq
$$
and so the second assumption is implied by the first unless the masses
are very disparate or $e^\prime$ is very large.  (This last case,
the ``impulsive limit'', is briefly considered in Section~3.2 and 
Appendix~A4.)

Now we develop the equations of motion.
Let $\br$ be the position of $m_2$ relative to $m_1$, and $\bR$ the
position of $m_3$ relative to the barycentre of the binary.  Then the
equation for the relative motion of the binary components is
$$
\ddot\br = {-GM_{12}\br\over r^3} + \bF,\eqno\stepeq
$$
with
$$
\bF = {Gm_3\over
R}\sum_{n=0}^\infty {m_1^{n-1} - (-m_2)^{n-1}\over
M_{12}^{n-1}}\grad_\br\left[\left({r\over R}\right)^n P_n\left({\br\cdot\bR\over
rR}\right)\right],\eqno\stepeq
$$
where $P_n$ is the $n$th Legendre polynomial.

Except for a factor $-m_1m_2/M_{12}$, equation~(4) is simply the gradient of the mutual potential of the
components of the binary and the third body, expressed as a multipole
expansion in powers of $r/R$.  The monopole and dipole terms
(i.e., those given by $n = 0$ and $1$) make no contribution to $\bF$,
and will play no further part in this paper.  

Our aim is to compute the change in eccentricity of the binary over
the entire duration of the encounter with the third body.  To do this
we have found it convenient to use the ``eccentric axis'' (Pollard
1976), which is a vector  also associated with the names of Laplace,
Hamilton, Lenz and Runge (Goldstein 1980).  For unperturbed Keplerian
motion it is a constant vector which points along the major axis of
the elliptic orbit of the binary components, and its magnitude is $e$.
Its definition is
$$
\be = \displaystyle{{1\over GM_{12}}}\dot\br\times(\br\times\dot\br) -
\displaystyle{{\br\over r}}.\eqno\stepeq
$$
Using  equation (3) it is easy to show that, in the presence of a
perturbing acceleration $\bF$, its variation is determined by
$$
GM_{12}\dot\be = 2(\bF\cdot{\dot\br})\br - (\br\cdot{\dot\br})\bF 
- (\bF\cdot\br)\dot\br\eqno\stepeq
$$ 
(cf.\ also Burdet 1967).
These equations are the basis of the theory described in Appendix A,
but some of the results we shall now summarise are more easily
understood with reference to these formulae.

\subsection{Binaries with Non-Zero Initial Eccentricity}

We begin with what would seem to be the most straightforward and
general case, that of a binary of arbitrary initial eccentricity $e$.
Also, since we are assuming that $\rp\gg a$, it is natural to neglect
all but the lowest non-vanishing term in equation~(4).  From what has been
said, this is the quadrupole term $n=2$ in the interaction between the
binary and the third body.

   From the assumption that the encounter is slow (\S2.1) it follows that,
to lowest order, perturbations may be
computed by first averaging over the fast relative motion of the
components of the binary. The resulting estimate of the change in
eccentricity is given (with a sign error) in the case of parabolic
motion of the third body by Heggie (1975, eq.~[5.66]), and it is
rederived in the general case (and with the correct sign) in 
Appendix~A to the present paper.

In order to state this result, we introduce standard angles describing
the relative orientation of the orbits.  The line of nodes is defined
to be the line of intersection of the orbital planes of the third body
and the binary.  Then the ascending node is the part of this line at
which the third body crosses the plane of the binary in the direction
of the angular momentum vector of the binary. 
Let $\Omega$ be the longitude
of the ascending node, measured in the plane of motion of the binary (in the sense
of the motion of its components) from the direction of pericentre of
the binary.  Let $i$ be the inclination of the two orbital planes, and
let $\omega$ be the longitude of pericentre of the third body,
measured in its plane of motion from the ascending node, in the sense
of its motion round the binary.
Then the result is that the
change in eccentricity is (cf.\ Appendix A1, eq.~[A4])
$$\eqalign{
\delta e = &-{15\over4}{m_3\over \sqrt{M_{12}M_{123}}}\left({a\over
\rp}\right)^{3/2}{e\sqrt{1-e^2}\over (1 + e^\prime)^{3/2}}~\times\cr
&\times\Bigg\{\sin^2i\sin2\Omega\left[\arccos(-1/e^\prime) + \sqrt{e^{\prime^2}-1}\right]
+ {1\over3}\left[(1+\cos^2i)\cos2\omega\sin2\Omega + 2\cos
i\sin2\omega\cos2\Omega\right] {(e^{\prime^2}-1)^{3/2}\over e^{\prime^2}}\Bigg\}.\cr
}\eqno\stepeq$$
This result is admittedly quite inelegant, but the result for parabolic motion
of the third body is neater; in that case
$$
\delta e = -{15\pi\over 16} \left({2 m_3^2 a^3\over M_{123} M_{12}
\rp^3}\right)^{1/2}e\sqrt{1-e^2}
\sin2\Omega\sin^2i.\eqno\stepeq
$$

The stated dependence on the ratio of the distances $a/\rp$ is easily
understood by examination of equations~(4) and~(6).  The quadrupole term
in equation~(4) is proportional to $a/\rp^3$, and so the terms in equation~(6)
are proportional to $a^3/(T\rp^3)$, where $T$ is the period of the
binary.  By Kepler's Third Law the duration of the encounter is of
order $T(\rp/a)^{3/2}$, and so the integration of equation~(6) with respect
to time gives a result proportional to $(a/\rp)^{3/2}$, as we have
found.

The dependence on $\Omega$ in equation~(8) is also not hard to
understand, at least for equal masses. 
Consider the case $\omega = 0$, $i = 90^\circ$, for example, and imagine
the orbit-averaged binary as a distribution of mass extended along its
major axis.  If $\Omega = 0$ or $90^\circ$ then the orbit of the third
body is situated on a plane of symmetry of the binary, and can exert no
net torque.  This already suggests a dependence on $\sin2\Omega$. 
Furthermore, if $\Omega = 45^\circ$, the torque {\sl increases} the angular
momentum of the binary, decreasing its eccentricity, which is consistent
with the sign of the right side of equation~(8).  On the other hand neither the
dependence on $i$ nor the independence on $\omega$ can be understood
from such considerations, as equation~(7) shows that the situation for
hyperbolic encounters is different.

\subsection{Initially Circular Binaries}

\subsubsection{Secular Octupole Perturbation: The Power-Law Regime}

Unfortunately this lowest-order result vanishes if the initial
eccentricity is zero.  This is evident from equation~(7) and can also be
understood from a rather general point of view, as discussed in
Section~2.3.3 and Appendix~B.  For a non-trivial result in this case it is
necessary to include the octupole
contribution to the interaction, i.e., the term with $n=3$ in equation~(4).
Details of the calculation are given in Appendix~A2, and
the result for the eccentricity induced in an initially circular binary
may be stated as follows:
$$\eqalign{
\delta e = {15\over 8}{ m_3\vert m_1 - m_2\vert\over M_{12}^2}
\left({M_{12}\over M_{123}}\right)^{1/2} &\left({a\over \rp}\right)^{5/2} 
{1\over e^{\prime^3}(1+e^\prime)^{5/2}} \times\cr
\times\bigg\{\cos^2i\sin^2\omega&\left[f_1(e^\prime)(1 - {15\over4}\sin^2i) +
{2\over15}(e^{\prime^2}-1)^{5/2}(1-5\sin^2\omega\sin^2i)\right]^2 +\cr
+ \cos^2\omega&\left[f_1(e^\prime)(1 - {5\over4}\sin^2i) +
{2\over15}(e^{\prime^2}-1)^{5/2}(1-5\sin^2\omega\sin^2i)\right]^2\bigg\}^{1/2},\cr
}\eqno\stepeq
$$
where
$$
f_1(e^\prime) = e^{\prime^4}\arccos(-1/e^\prime) +
{\sqrt{e^{\prime^2} - 1}\over15}(-2 + 9e^{\prime^2} + 8e^{\prime^4}).\eqno\stepeq
$$

Again the result for parabolic motion of the third body is less
inelegant.  In fact for this case we have
$$
\delta e = {15\pi\over 32} {m_3\vert m_1 - m_2\vert\over 
M_{12}^2}\left({M_{12}\over 2M_{123}}\right)^{1/2}\left({a\over \rp}\right)^{5/2}
\left[\cos^2i\sin^2\omega(1-{15\over4}\sin^2i)^2 +
\cos^2\omega(1-{5\over4}\sin^2i)^2\right]^{1/2}.\eqno\stepeq
$$

As before, it is easy to understand the dependence of these results on
$a/\rp$. The absence of $\Omega$ in these results should also come as
no surprise, since the binary is assumed to be circular, and we have
averaged over the position of its components.  Another point of
interest is the fact that the results are proportional to $\vert m_1 -
m_2\vert$, and the reason for this is easily seen by examination of
the $n=3$ (octupole) term in equation~(4).

In general, for a circular binary with unequal masses, the result~(9) is the dominant
contribution to the induced eccentricity at sufficiently
large $\rp$. We refer to this as the ``power-law regime'', since the change in
eccentricity is still proportional to a power of $\rp/a$, although it is a
higher power than in the case of non-zero initial eccentricity. In contrast,
at smaller $\rp$, or in the special case of equal masses, we show in the next section
that the  induced eccentricity drops exponentially with $\rp$.

\subsubsection{Non-Secular Quadrupole Perturbation: The Exponential Regime}

In Section 2.2  we showed that, if we average over the motion of the binary
and take only the quadrupole interaction, then the eccentricity
induced in an initially circular binary vanishes.  At small $\rp$,
however, the assumption on which the method of averaging is based
breaks down, and here we re-examine the problem without averaging.
The result turns out to be important in applications.  The derivation
is deferred to Appendix~A3, where it is shown that
$$\eqalign{
\delta e = 3\sqrt{2\pi} {m_3 M_{12}^{1/4}\over M_{123}^{5/4}}
    \left({\rp\over a}\right)^{3/4} {(e^\prime +1)^{3/4}\over e^{\prime^2}}
&\exp\left[-\left({M_{12}\over M_{123}}\right)^{1/2}\left({\rp\over a}\right)^{3/2}
{\sqrt{e^{\prime^2}-1}-\arccos(1/e^\prime)\over(e^\prime-1)^{3/2}}\right]\times\cr
&\times\cos^2{i\over2}\left[\cos^4{i\over2} + {4\over9}\sin^4{i\over2} +
{4\over3}\cos^2{i\over2}\sin^2{i\over2}\cos(4\omega + 2\Omega)\right]^{1/2}.\cr
}\eqno\stepeq
$$
Here $\Omega$ is defined somewhat differently than in Section~2.2: it is
measured in the plane of motion of the binary, from its position at the
time of closest approach of the third body.  Therefore this
definition, and this formula, includes a dependence on the 
phase of the binary. Again this result 
simplifies a little for parabolic motion of the third body, to 
$$\eqalign{
\delta e = 3\sqrt{2\pi} {m_3 M_{12}^{1/4}\over M_{123}^{5/4}}
\left({2\rp\over a}\right)^{3/4}
&\exp\left[-{2\over3}\left({2M_{12}\over
M_{123}}\right)^{1/2}\left({\rp\over a}\right)^{3/2}\right]\times\cr
&\times\cos^2{i\over2}\left[\cos^4{i\over2}
 + {4\over9}\sin^4{i\over2} +
{4\over3}\cos^2{i\over2}\sin^2{i\over2}\cos(4\omega + 2\Omega)\right]^{1/2}.\cr
}\eqno\stepeq
$$

The approximations used in the derivation of these formulae again make use of
the assumption that $\rp\gg a$.  Indeed for sufficiently large $\rp$
these results are negligible in comparison with equations~(7) and (8) if the
binary is initially non-circular, or with equations~(9) and (11)
if $e = 0$ initially.  Nevertheless it turns out to be an important
pair of results, which are dominant in a small but significant range of
near-encounter distances where the exponential term is not so small
as to be masked by the previous results.  

There is one situation in which equations~(12) and (13) appear to dominate at {\sl
all\/} $\rp\gg a$:   the case of a circular binary with equal
masses.  This is not a rigorous result, but we do show in Appendix~B
that the induced eccentricity vanishes with increasing $\rp$ more
quickly than any power of $a/\rp$.  We already remarked that the
octupole results, equations~(9) and (11), are proportional to the mass difference
$\vert m_1 - m_2\vert$.

Looking at equations~(12) and (13) one is tempted to conclude that the eccentricity is
a kind of adiabatic invariant.  It is usual, however, to apply this
term to a dynamical quantity whose variation is small when the system
is subject to a large but slow perturbation.  Here the perturbation is
not only slow but small.  Furthermore, the exponential form of equations~(12)
and (13)
only applies to a tiny set of orbits in which the initial eccentricity
vanishes, whereas adiabatic invariants enjoy their special properties
more globally.  Henceforth we refer to this result simply as ``the
exponential formula'', and when it is dominant we say that we are in
the ``exponential regime''.

It is not as easy as in the previous cases to understand the 
dependence of these results on $a/\rp$, except for the exponential
term.  The exponent is simply of order the ratio of the two main
timescales in the problem: the period of the binary and the duration
of the encounter.  Such a term arises in other well studied problems
of this kind (cf.\ Goldstein 1980, eq.~[11-138]).

Finally, although the results in equations~(12) and (13) vanish in the
retrograde case ($i = 180^\circ$), this simply means that the leading
term in this case is of a higher order in $a/\rp$, and we have neglected
all such terms in the derivation of these results.

\subsubsection{Remarks on First-Order versus Second-Order Perturbations}

Details of the derivation of the foregoing results are given in 
Appendix~A, but it is worth exposing here one or two analytical issues which
help to clarify the nature of the results.

There are several methods by which the computation of the change in
eccentricity might be attempted.  As already stated in Section~2.1, 
we have used the eccentric axis
(Pollard 1976).  Other possible methods are the use of Lagrange's planetary equations,
though in their usual form they are unsuited for computations of
low-eccentricity orbits, and it is better to express them in terms of
Poincar\'e variables (Plummer 1918).  From a physical point of view the
most obvious method is to compute the change in angular momentum.  This
involves the changes in both $e$ and $a$, but the latter is already
known (Heggie 1975, eq.~[5.40]), and so the change in $e$ alone may be
determined.

Whichever approach is adopted, the next step is a method of successive
approximations. At lowest (first) order, unperturbed Keplerian motions are
substituted into the perturbation terms.  At next order, the first-order
results are substituted, and so on.  In addition, it is possible to
express the perturbation term itself as a multipole expansion.

As we have seen, if we work to lowest order, retain only the
quadrupole interaction, and average over the fast motion of the binary,
then the eccentricity induced in a circular binary
vanishes.  It is tempting to offer the following
explanation. After averaging, the binary components are effectively
replaced by two rings of matter, which coincide if $m_1 = m_2$.  The
density of each ring is proportional to the time spent by a component at
each point. For a circular binary this ring is uniform and circular, and
so the force field experienced by the third body is axisymmetric.
Therefore the axial component of angular momentum of the third body
(i.e., the component perpendicular to the plane of motion of the binary
components) is conserved, and it follows, from overall angular momentum
conservation, that the axial component of angular momentum of the binary
is also conserved. It follows that the change in the magnitude of the
angular momentum of the binary also vanishes (to first order). Since
the change in semi-major axis over the entire encounter is exponentially
small (Heggie 1975), 
it follows that the change in eccentricity also
vanishes to first order.  Note that this argument applies even without
an expansion in powers of $a/\rp$;  it depends only on the use of
averaging and the use of first-order perturbation theory (whereby it is
assumed that the interaction is calculated as if the three bodies
proceed on their unperturbed orbits.)  Therefore it also follows that
the octupole contribution would vanish to first order, and indeed all terms of
the multipole expansion.

At second order it is not so clear what to expect.  Averaging at first
order ignores the fact that the third body causes the eccentricity
of the binary to oscillate, and when this is taken into account (at
second order) it is
unclear whether the averaged mass distribution of the components is
still axisymmetric.  In their interpretation of their numerical results,
Hut \& Paczy\'nski (1984) considered that the change in eccentricity must
indeed be a second-order effect.  We have computed the
second-order change in the eccentricity (via the change in the angular
momentum), and find a non-zero result in general in the case $m_1\neq
m_2$.  This result is generated by the octupole term.

It turns out, however, that the distinction between first- and
second-order effects is not an absolute one, but depends on the
variables used in the computation.  In the present paper we adopt the
Lenz vector in preference to the angular momentum, as a means of
computing the eccentricity.  With this variable it turns out that a
non-trivial result for the eccentricity induced in a circular binary
is obtained at {\sl first} order (provided again that $m_1\neq m_2$.)  The
result agrees with that obtained from a second-order computation of
the change in angular momentum, and is also caused by the octupole
interaction, but it is much more easily calculated using the Lenz
vector, as only first-order perturbation theory is needed.  It is not
hard to see why second-order perturbation theory is required if the
angular momentum is used.  The angular momentum is proportional to
$\sqrt{1-e^2}$, and so if the initial eccentricity is zero, the change in
angular momentum is second order in the change of eccentricity.

Finally in this section we turn to the case of a binary with equal
masses.  Here the odd-order terms in the multipole expansion, equation~(4),
 vanish, and it is easy to see that all first-order
calculations, whether with the Lenz vector or the angular momentum, will
yield a null result.  Consider, for example, the method adopted in
Appendix~A.  The perturbation is an odd function of $\br$ (the position
vector of one component relative to the other), and it follows from
equation~(6) for the rate of change of the eccentric
axis that this is also odd.  A null result is obtained, therefore, if
we average over the motion of the components.  This result, which is so
simple to prove at first order, also applies at all orders of
perturbation theory, but the proof is somewhat technical, and is
relegated to Appendix~B. The implication of the result, however, is that
the change in eccentricity must vanish more quickly (with increasing
$\rp$) than any power of the expansion parameter $a/\rp$, and it is
likely that the result vanishes exponentially as an inverse power of
this parameter (cf.\ Section 2.3.2).   

The problem of {\sl rigorously} justifying the computation of
exponentially small perturbations is one which has attracted
considerable attention from mathematicians in recent years (e.g.,
Byatt-Smith \& Davie 1990). The difficulty is, of course, that standard
perturbation methods proceed by expanding in powers of a small quantity,
whereas these exponentially small results have no power series
expansion.  Unfortunately it still appears that each case must be
treated separately if it is to be handled rigorously, and there is no
general calculus for this class of problems. The present problem is
certainly considerably more elaborate than those which have been treated
properly.  Our interest in this result is more pragmatic, however.  We
shall see that the result of Section~2.3.2, which exhibits exponential
dependence, is important when it dominates the result from the octupole
interaction (Section~2.3.1), and in this situation it does not require elaborate
justification.

\subsection{Illustrations and Comparison with Numerical Results}

We have performed a large number of numerical integrations of encounters
between a binary and a third star. In this section 
we present a sample of results, to establish
the accuracy of our analytical expressions and to illustrate the results for
some typical cases representative of the different possible regimes. 

We shall focus primarily on parabolic encounters, which are particularly relevant
astrophysically since the conditions in globular clusters are very close
to the parabolic limit except for very wide binaries ($a\gg 1\,$AU). In
general, in any star cluster, hyperbolic encounters are of importance
only for binaries that are sufficiently wide to be classified as
``soft'' and therefore will tend to be easily disrupted by
interactions (Hills 1975; Heggie 1975; Hut 1983). Deviations from the
parabolic results remain small as long as the relative velocity at
infinity for the encounter is less than the orbital velocity of the
binary, and the encounter is not too distant (cf.\ eq.~[1] and Section~3.2 below).

\subsubsection{Numerical Integrations}

The numerical integrations are fairly straightforward.
We have used the Bulirsch-Stoer algorithm described in Press et al.\ (1992) to
integrate the differential equations of motion of the three-body problem directly.
Total energy and total angular momentum conservation is maintained typically to within
$10^{-12}-10^{-11}$, for computations in double precision. This allows changes
in eccentricity as small as $\delta e\sim 10^{-8}$ to be measured accurately.
The input parameters for a given integration are the masses $m_1$, $m_2$
and $m_3$, the semi-major axis $a$, eccentricity $e$, and mean anomaly $\phi_i$
(or orbital phase, measured from pericentre) of the binary at $t=0$, 
the periastron distance $\rp$ and velocity
at infinity $V$ characterizing the outer orbit (cf.\ Section~2.1), 
the initial separation $r_i$
(distance between $m_3$ and the centre of mass of the binary at $t=0$), and
the angles $\Omega$, $i$, and $\omega$ specifying the relative orientation of
the two orbits (defined in Section~2.2). We generally adopt units such that
$G=m_3=a=1$. 
Each integration follows the system through periastron passage and is continued
until the final separation has increased to a value equal to the initial
separation, $r_f=r_i$. Typically we use $r_i/\rp\sim10^2$. For a given set of
input parameter values, we normally repeat the integrations for many different
values of the initial phase $\phi_i$, covering systematically the interval
from 0 to $2\pi$. This allows us to compute a phase-averaged result and to
study the extent of the phase dependence.

\subsubsection{Dependence on the Periastron Distance}

First we consider three simple situations where one of the three expressions~(8),
(11), and~(13) dominates over the entire range of periastron separations where
$\delta e$ remains small enough that a perturbative calculation applies.

In Fig.~1, we show the change in eccentricity of a binary with non-zero
initial eccentricity, following an encounter with a star on a general
inclined orbit. The three masses are assumed identical. The result is
given by equation~(8) to high accuracy. Here and in all subsequent figures
the solid lines show the analytical results and the dots show the numerical results,
with the ``error bars'' indicating the full extent of the dependence on 
orbital phase. In this case the phase dependence remains completely negligible
down to very small values of $\rp/a$. Notice that, as long as the three masses
are comparable, the magnitude of the fractional change in eccentricity $|\delta e|/e$
predicted by equation~(8) is always very small. Using equation~(8) we see that 
a change $|\delta e|/e\ga1$ would require $m_3 \ga (\rp/a)^3 M_{12}$.

\beginfigure{1}
\epsfxsize 4in
\centerline{
\epsffile{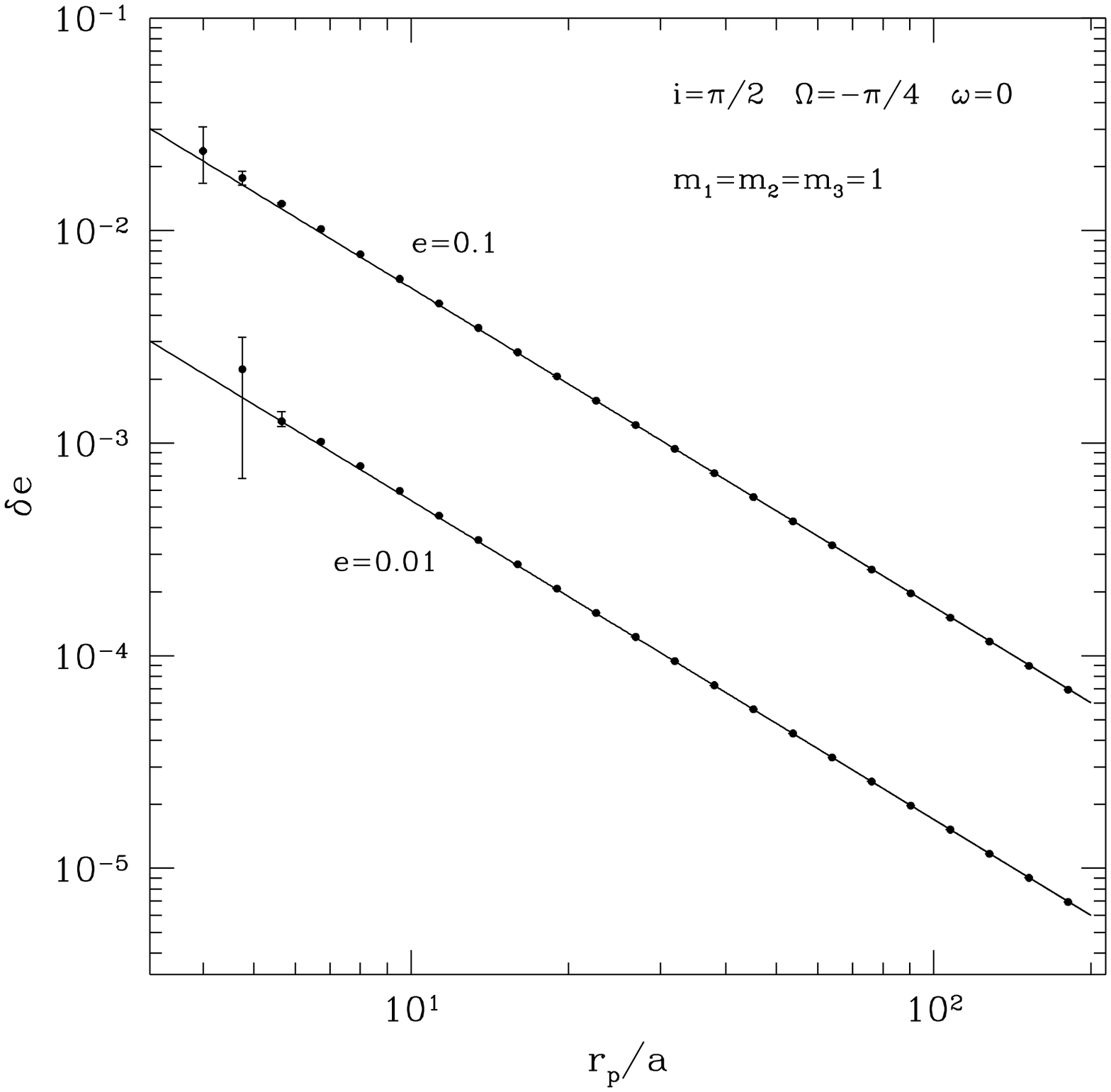}
}
\caption{{\bf Figure 1.} 
Change in eccentricity $\delta e$ for an initially non-circular binary 
(initial eccentricity $e=0.1$ or 0.01) perturbed by
an encounter with a third star on an inclined, parabolic orbit with
periastron distance $\rp$ (given in units of the binary semi-major axis $a$). 
The dots show the results of numerical integrations; the lines show the analytical 
power-law result, equation~(8). The ``error bars'' indicate the full extent of 
the phase dependence. For $\rp/a\ga5$ this phase dependence is no longer 
visible.}
\endfigure

Now we turn to initially circular binaries.
To isolate the octupole effect (equation~[11]), we consider a binary with $m_1\ne m_2$ and
we focus on distant encounters, with $\rp/a\ga5$. The variation of the
induced eccentricity $\delta e$
with $\rp$ is shown in Fig.~2 for a coplanar, prograde encounter, and for two
different values of $m_2$. Notice the steeper power-law behavior ($\rp^{-5/2}$)
compared to the case of an initially non-circular binary ($\rp^{-3/2}$). 
Notice also the large drop in $\delta e$ when $m_2$ becomes nearly equal to $m_1$.
To best illustrate the exponential regime (equation~[13]), 
we show in Fig.~3 the variation of $\delta e$ in the equal-mass case. 
The induced eccentricity now drops much more rapidly, to
values $\ll 10^{-7}$ for $\rp/a\ga 10$.
Notice that equation~(13) continues to predict 
$\delta e$ with remarkable accuracy down to very small values of $\rp/a$.

\beginfigure{2}
\epsfxsize 4in
\centerline{
\epsffile{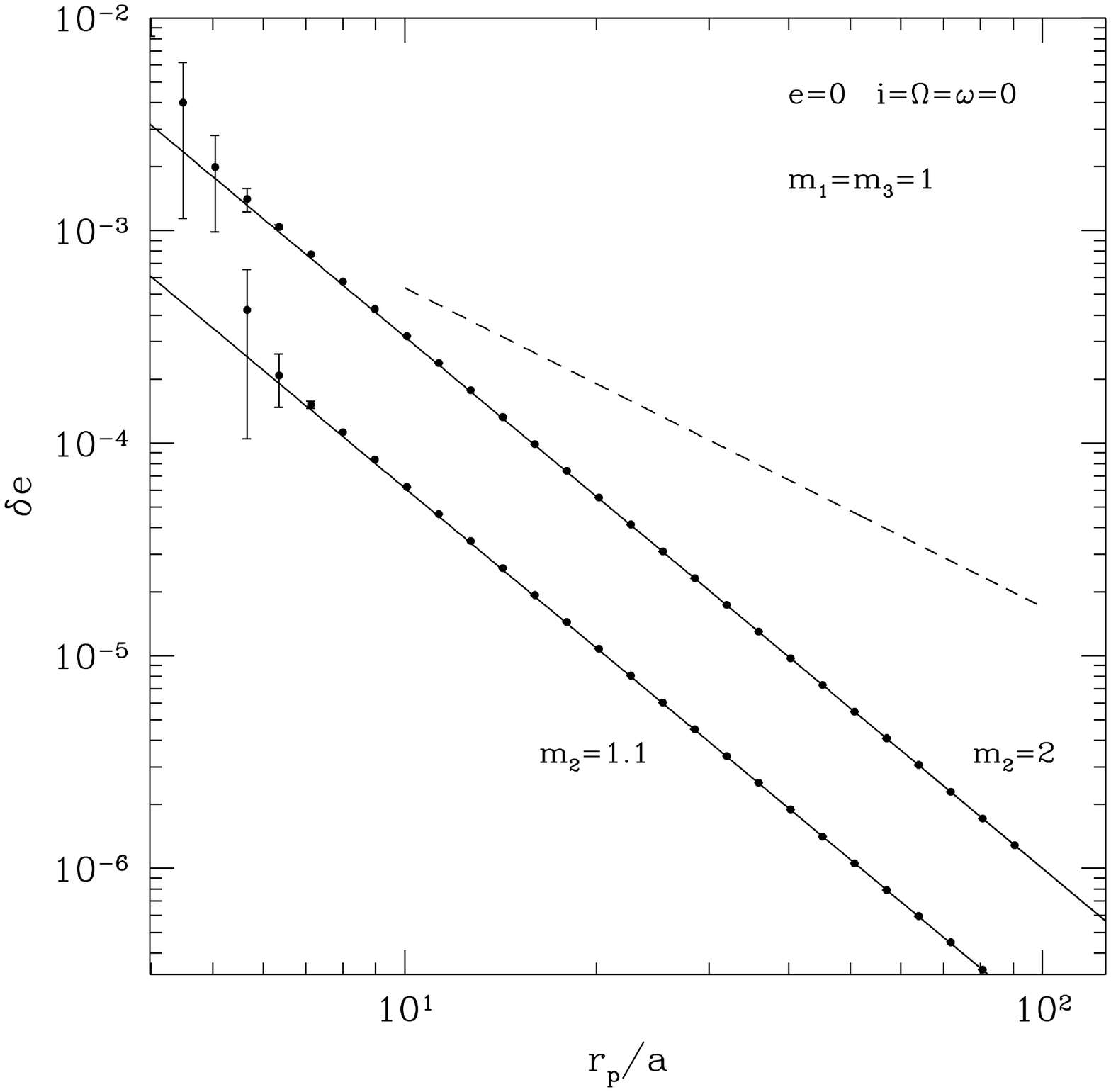}
}
\caption{{\bf Figure 2.} 
Change in eccentricity $\delta e$ for a binary with initial $e=0$ 
perturbed by
an encounter with a third star on a coplanar, prograde, parabolic orbit with
periastron distance $\rp$ (given in units of the binary semi-major axis $a$). 
Conventions are as in Fig.~1. The lines show the analytical result, equation~(11),
for two different values of $m_2$.
The dashed line shows, for comparison, the
variation of $\delta e$ for a non-circular binary with initial $e=0.01$ (same
case as in Fig.~1).}
\endfigure

\beginfigure{3}
\epsfxsize 4in
\centerline{
\epsffile{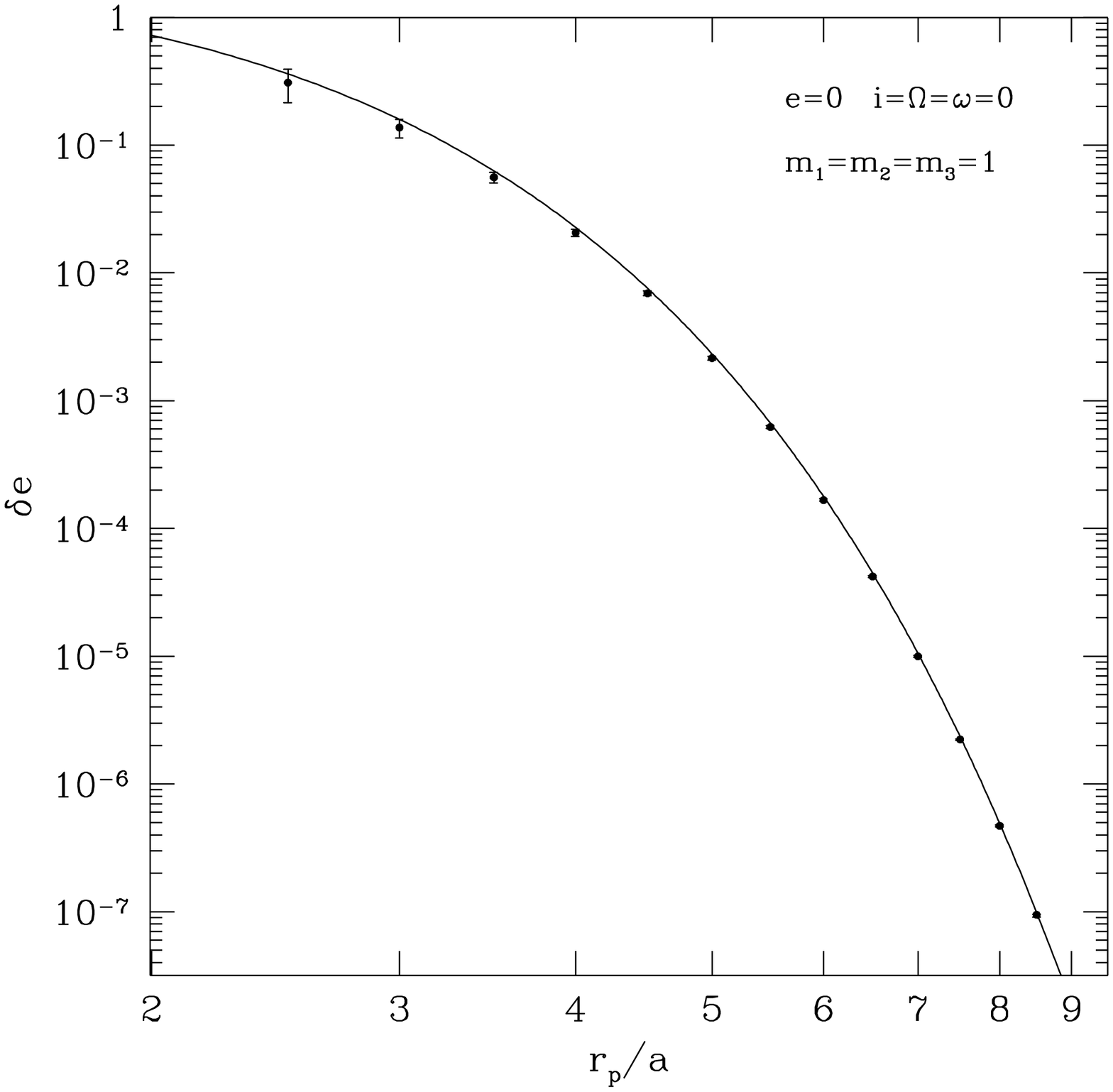}
}
\caption{{\bf Figure 3.} 
Change in eccentricity $\delta e$ for a binary with initial $e=0$ 
and containing two identical masses $m_1=m_2$.
Conventions are as in Figs.~1 and~2. The line shows the analytical 
result, equation~(13).}
\endfigure

In a general case, all three types of perturbations dominate in different
ranges of $\rp$. Consider for example a typical binary millisecond pulsar
containing a neutron star of mass $m_1=1.4\,M_\odot$ with a low-mass companion
$m_2=0.2\,M_\odot$ in a nearly circular orbit of initial eccentricity $e=10^{-6}$;
the binary is in the dense core of a globular cluster where the average stellar mass is
$m_3=1\,M_\odot$ (cf.\ Rasio \& Heggie 1995). Fig.~4 shows the final eccentricity
$e_f=e+\delta e$ following a coplanar, prograde encounter, as a function 
of $\rp/a$. The exponential regime corresponds to $\rp/a\la5$. For $\rp/a\ga200$,
equation~(8) dominates and the fractional change $\delta e/e$ is very small.
In a wide range of intermediate values, $5\la \rp/a \la 200$, 
the octupole power-law result is dominant. As discussed in Rasio \& Heggie (1995),
an important consequence of these results is that the cross section for inducing
eccentricity in binary millisecond pulsars in clusters is much larger than was
estimated in previous studies (see also section 3 below).

\beginfigure{4}
\epsfxsize 4in
\centerline{
\epsffile{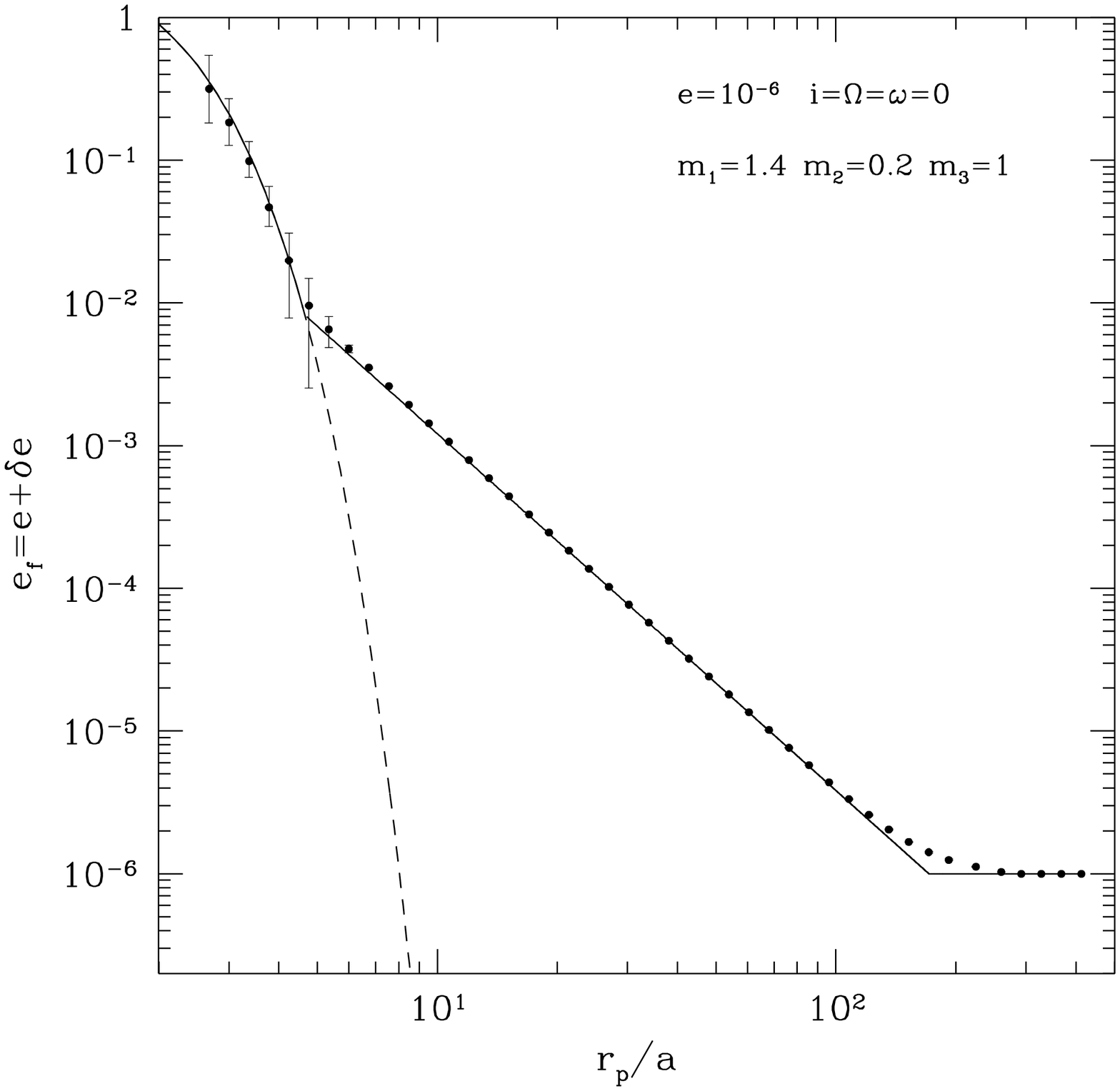}
}
\caption{ {\bf Figure 4.} 
Final eccentricity $e_f=e+\delta e$ for a binary with initial $e=10^{-6}$ 
following a coplanar, prograde encounter. Conventions are as in Fig.~1.
The masses have values representative of a typical binary millisecond
pulsar in a globular cluster. 
The solid lines show the three separate analytical results, equations~(8), (11),
and~(13). The dashed line shows the continuation of the exponential result,
equation~(13), for comparison. Notice 
that the octupole power-law result,
equation~(11), dominates over a wide range of intermediate values of $\rp$.
}
\endfigure

\subsubsection{Angular and Phase Dependences}

Although we have checked very carefully most of the phase and angular dependences
appearing in our analytical expressions, we shall only present a small number
of illustrative cases here. We focus on initially circular binaries, since
the angular dependent factors appearing in equation~(8), for the case of
non-zero initial eccentricity, can be understood from elementary
arguments, at least in part. 
We use the same typical values of the masses for a binary millisecond pulsar in a
globular cluster as in Fig.~4, and we consider only parabolic encounters.

First, we look at the dependence on inclination
for the induced eccentricity in the power-law regime. Fig.~5 shows the
results for $\rp/a=30$ and for two particular values of $\omega$. For $\omega=0$,
the first term in the angular dependent factor of equation~(11) vanishes and
the dependence on inclination is $\propto\vert 1 - (5/4)\sin^2 i \vert$,
giving zeros of $\delta e$ at $i\simeq0.352\pi$ and $i\simeq0.648\pi$.
The maximum $\delta e$ is obtained for $i=0$ and $i=\pi$, with a secondary
maximum at $i=\pi/2$.
For $\omega=\pi/2$, the second term in the angular dependent factor vanishes
and $\delta e\propto\vert \cos i\, [1 - (15/4)\sin^2 i]\,\vert$, giving zeros at
$i=\pi/2$, $i\simeq0.173\pi$, and $i\simeq0.827\pi$. 
Again the maximum $\delta e$ is obtained for $i=0$ and $i=\pi$, with two secondary
maxima at $i\simeq0.335\pi$ and $i\simeq0.665\pi$. The agreement between
the numerical and analytical results is everywhere excellent, with deviations 
remaining always $\la3\%$.

In the exponential regime, a much stronger phase-dependence is observed, and
larger deviations are observed between the numerical and analytical results.
This is not surprising, since the exponential regime corresponds to rather
large values of the perturbation expansion parameter $a/\rp$. In Fig.~6, we
illustrate the strong phase-dependence as well as the dependence on inclination
for $\rp/a=3.5$. The phase-dependence is largest for $i=0$ (coplanar prograde
encounters), which also corresponds to the maximum value of $\delta e$
according to equation~(13). 
For $i=\pi$, equation~(13) predicts $\delta e=0$ even though the
numerical result is clearly non-zero. The small discrepancies between numerical
and analytical results in this regime come from neglecting
the more rapidly oscillating term in the derivation of equation~(13) (cf.\
Appendix~A3), and from the octupole perturbations.

\beginfigure{5}
\epsfxsize 4in
\centerline{
\epsffile{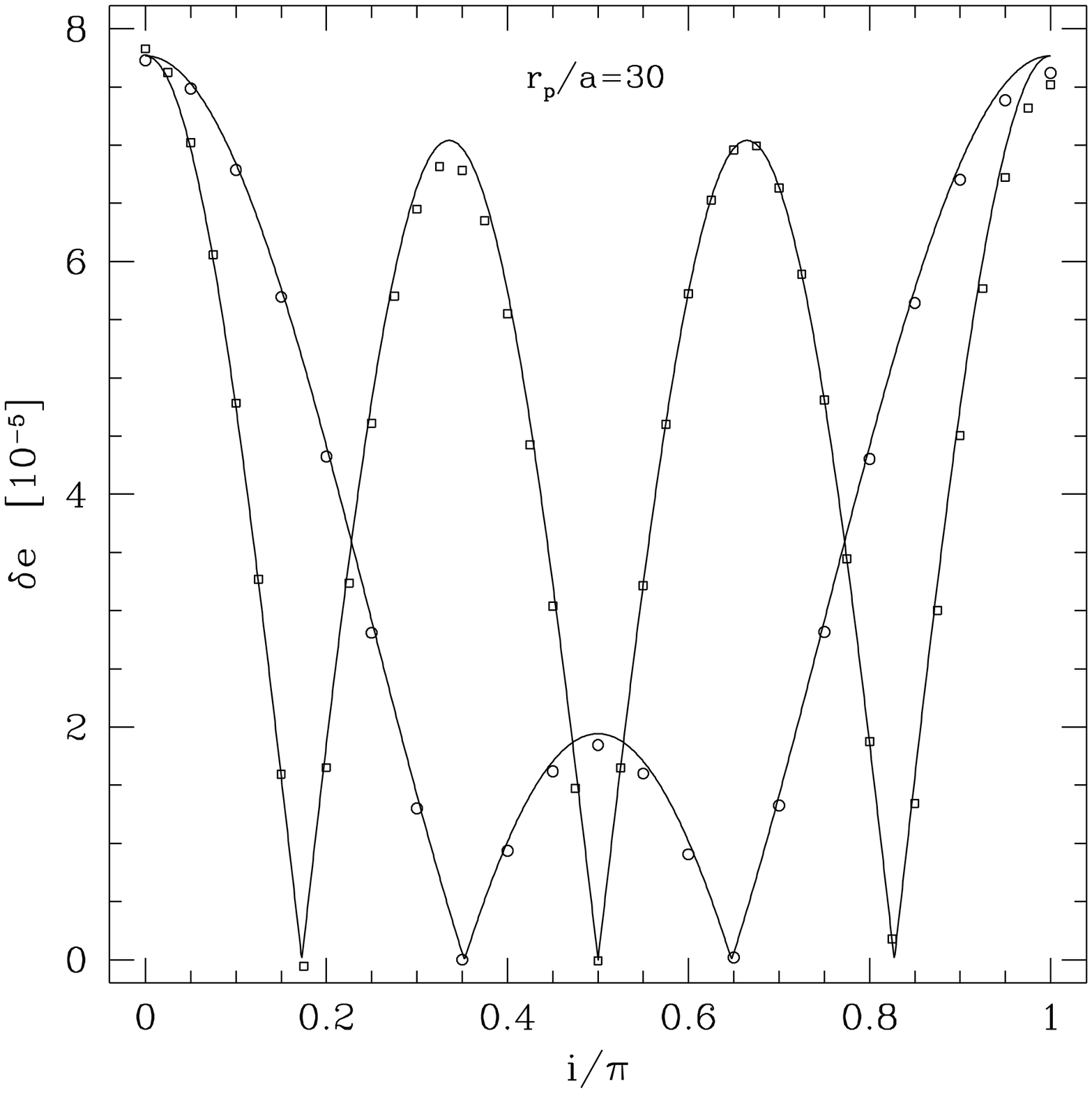}
}
\caption{ {\bf Figure 5.} 
Dependence of induced eccentricity $\delta e$ on inclination for an initially
circular binary with $m_1=1.4$, $m_2=0.2$, $m_3=1$, for a parabolic encounter with
$\rp/a=30$ (power-law regime). The round dots are for $\omega=0$, the square dots
for $\omega=\pi/2$. The lines show the analytical results, equation~(11). 
}
\endfigure

\beginfigure{6}
\epsfxsize 4in
\centerline{
\epsffile{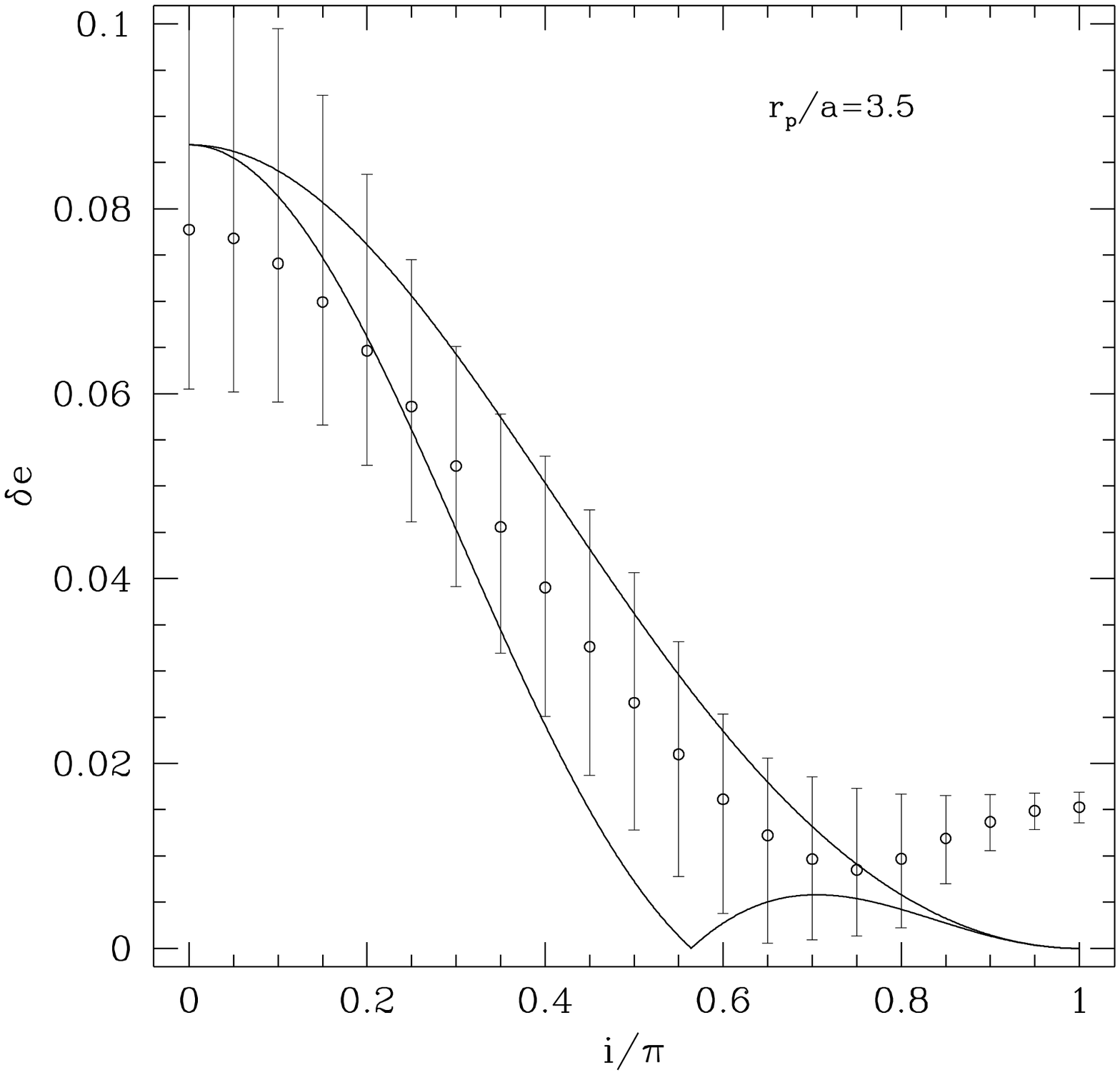}
}
\caption{{\bf Figure 6.} 
Same as Fig.~5 but for $\rp/a=3.5$ (exponential regime).
The two lines correspond to $\cos(4\omega+2\Omega)=\pm1$ in equation~(13),
i.e., they show the extent of the theoretically predicted phase dependence.
As before, the ``error bars'' show the full extent of the phase dependence
in the numerical results while the dots show the phase-averaged numerical results.
}
\endfigure

Other circumstances where a strong phase-dependence is observed are when
two different perturbation effects contribute similar-amplitude changes in the
eccentricity, as in Fig.~4 near $r_p/a\simeq5$, or when $\vert\delta
e\vert\simeq e$.  As an example of the latter consider the case of a
binary with $m_1=m_2$ and $i=0$ (so that both power-law
expressions [8] and [11] vanish) and with
a very small initial eccentricity $e=10^{-4}$, as illustrated in Fig.~7.
The principal contribution to $\delta e$ is given by equation (13),
though this was calculated for the case of an initially circular orbit.
Since in the present case $e$ is small but non-zero, $\delta e$ must be
interpreted as the magnitude of the vector $\delta\be$, and the final
eccentricity is given by $e_f = \vert\be + \delta\be\vert$.  Now the
direction of $\delta\be$ depends on the phase of the binary, as can be
seen by careful inspection of eq.(A21), and so if $\vert\delta\be\vert =
e$ we see that $e_f = 0$ for certain phases.  In fact eq.(13) shows that
the condition  $\vert\delta\be\vert = e$ is satisfied when
$r_p/a\simeq6.2$.  At this value the numerical result in Fig.~7 would have an ``error bar'' extending to arbitrarily
small values of $e_f$ in this logarithmic plot.  An alternative argument
may be constructed by considering the time-reversal of such an
encounter, i.e. one in which an initially circular binary is perturbed
into an orbit with final eccentricity $e_f=10^{-4}$.  Eq.(13) shows that
this can only occur at the same critical value of $r_p/a$.

\beginfigure{7}
\epsfxsize 4in
\centerline{
\epsffile{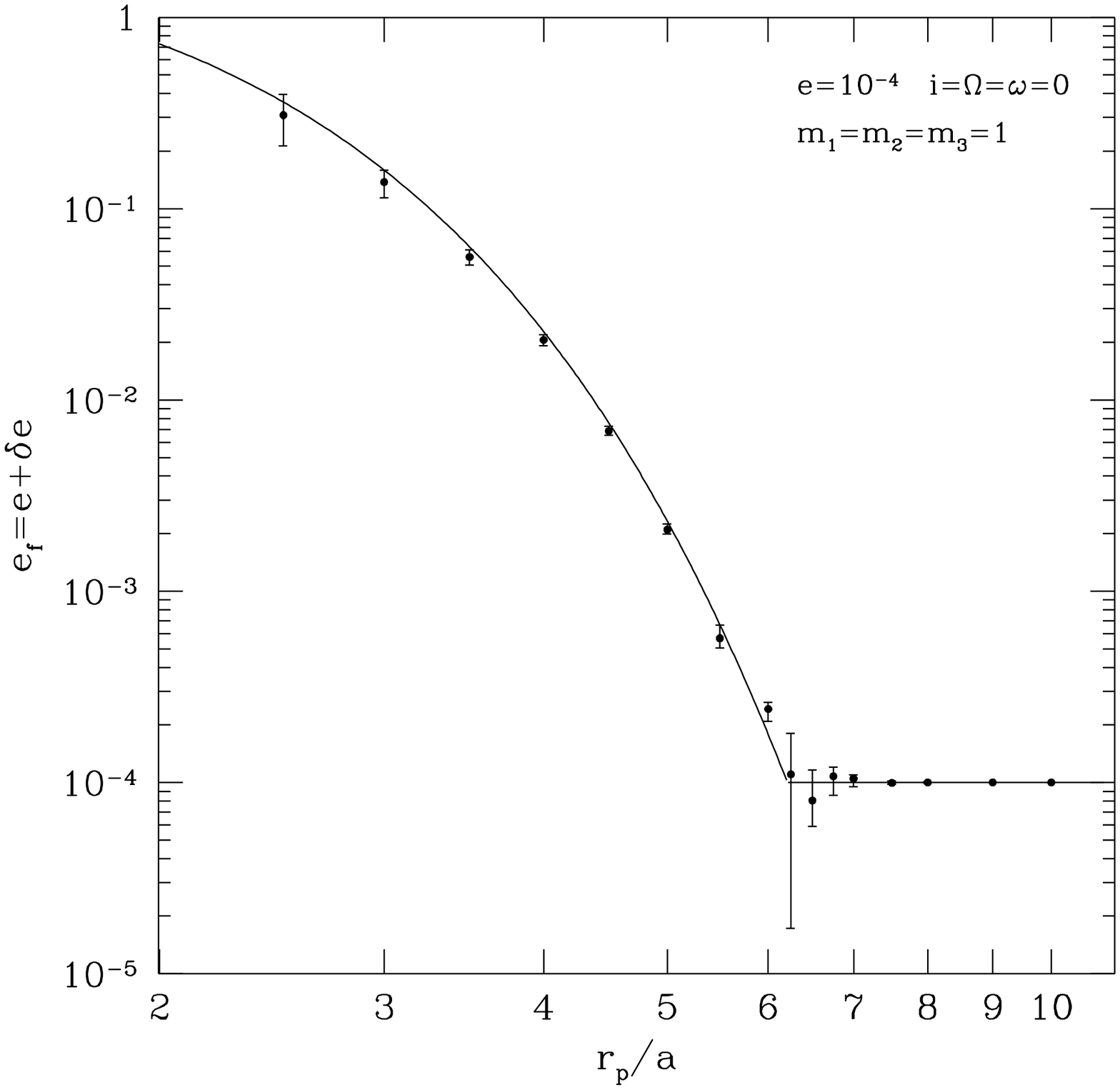}
}
\caption{{\bf Figure 7.} 
Same as Fig.~3 but for a binary with non-zero initial $e=10^{-4}$.
Notice the strong phase dependence near $\rp/a\simeq6-7$, where the
magnitude of $\delta\be$ (given by eq.~[13]) is
comparable with the initial value of $e$.
}
\endfigure

\subsubsection{Dependence on the Masses}

In general, when the three masses are all of comparable magnitude,
and $m_1\ne m_2$, the mass-dependence of the induced eccentricity $\delta e$ 
is rather weak. Near $m_1=m_2$, however, the power-law expression, equation~(11),
 can become arbitrarily small. 
The peculiar dependence of $\delta e$ on $\vert m_1-m_2\vert$
in the power-law regime is illustrated in Fig.~8 for a typical case.
The agreement between analytical and numerical results is again excellent.

When $m_3\gg M_{12}$, as in the case of an encounter between a binary and a massive
black hole, it is possible to find that the exponential result dominates
even at rather large values of $\rp/a$. This is because the duration of the encounter
decreases as $m_3$ increases at fixed $\rp$. In Fig.~9 we show the dependence of
$\delta e $ on $m_3$ for fixed $\rp/a=30$. The exponential result becomes dominant
when $m_3\ga300$. For $m_3\ga 3000$, the binary is disrupted by the encounter.

\beginfigure{8}
\epsfxsize 4in
\centerline{
\epsffile{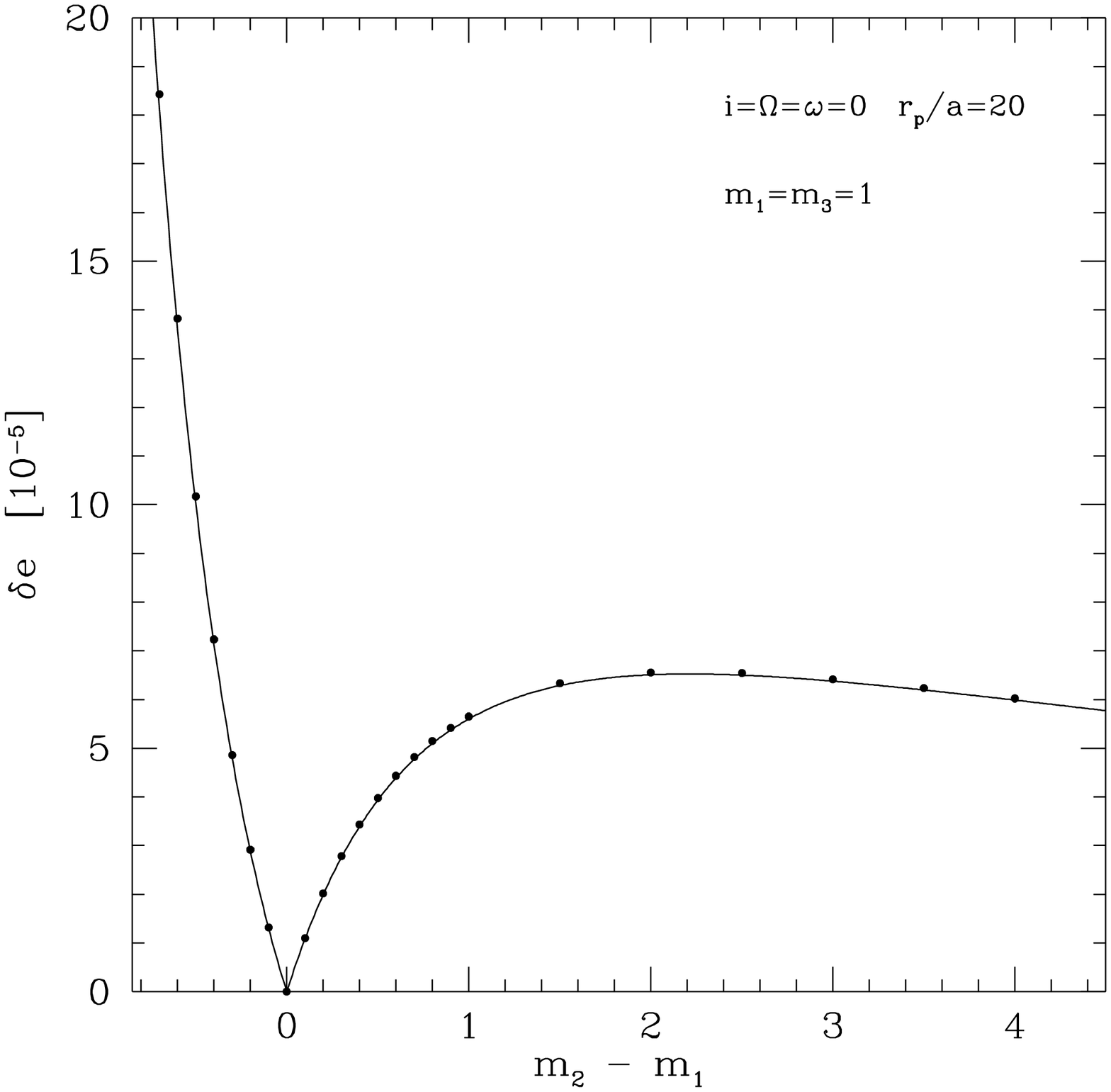}
}
\caption{{\bf Figure 8.} 
Dependence of $\delta e$ on the masses near $m_1=m_2$ for the power-law regime.
Equation~(11) predicts $\delta e=0$ for $m_1=m_2$, in agreement with the
numerical results.
}
\endfigure

\beginfigure{9}
\epsfxsize 4in
\centerline{
\epsffile{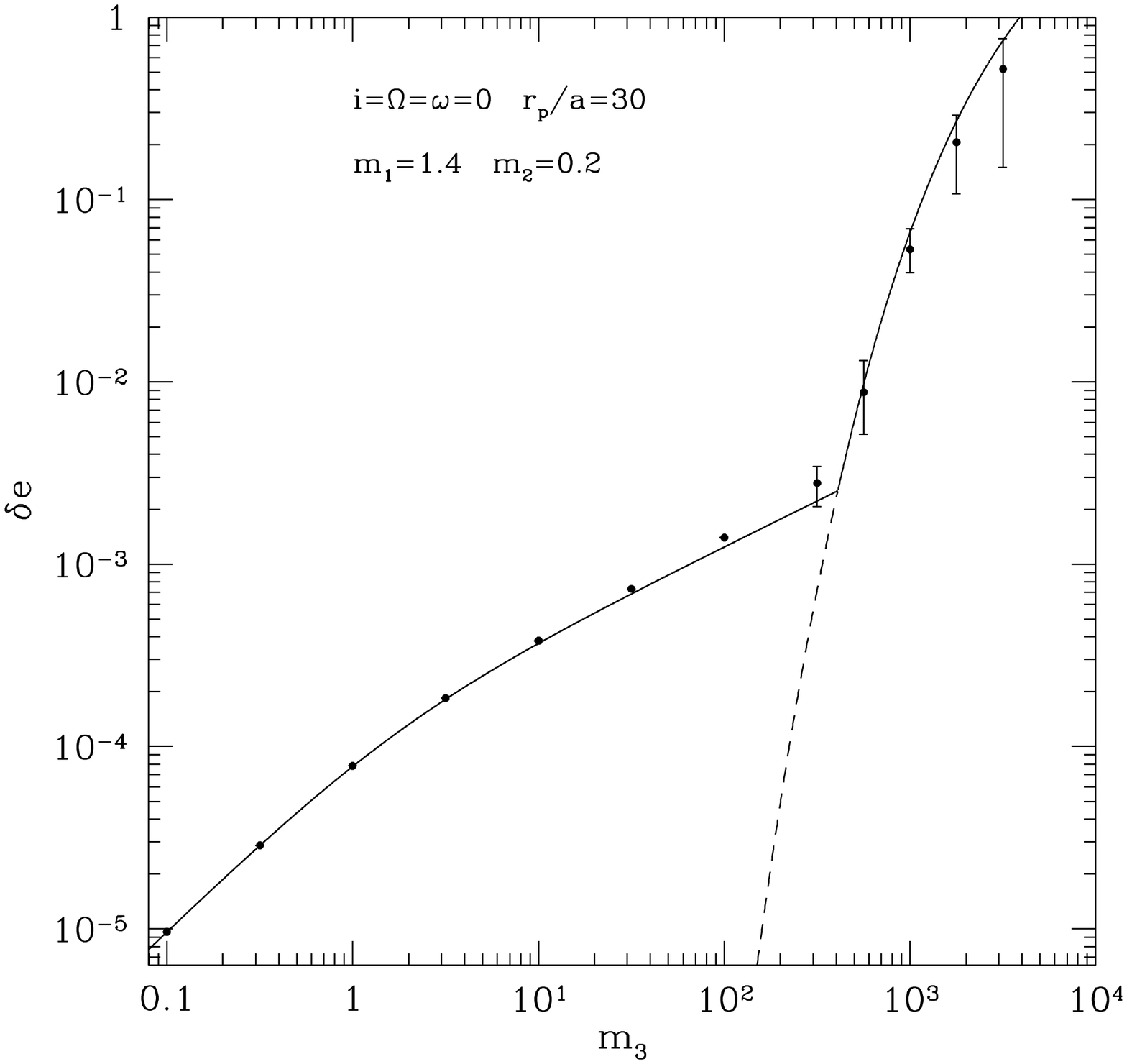}
}
\caption{{\bf Figure 9.} 
Same as Fig.~4 but showing the dependence of $\delta e$ on $m_3$. For sufficiently
large $m_3$ (here for $m_3\ga300$), non-secular effects can always become
dominant, even for encounters with large $\rp/a$.
}
\endfigure

\subsubsection{Hyperbolic Encounters}

For fixed $\rp$, as the relative velocity at infinity $V$ increases from zero, the 
duration of the encounter becomes shorter and shorter and we always expect the
exponential result, equation~(12), to dominate eventually. An example is shown
in Fig.~10, where we fix $\rp/a=30$. Although the secular octupole effect is
completely dominant in the parabolic limit, non-secular effects become dominant
for $V/(Gm_3/a)^{1/2}\ga4$, or $V\ga 120\,{\rm km}\,{\rm s}^{-1} (m_3/1\,M_\odot)^{1/2}
(a/1\,{\rm AU})^{-1/2}$. Between $V=0$ and this value, $\delta e$ {\sl decreases\/} by
almost an order of magnitude, but then increases again as $V$ increases in the exponential 
regime. In contrast, Fig.~11 shows a case where the exponential result dominates everywhere,
which typically happens at sufficiently small $\rp$. Here $\delta e$ {\sl increases\/} at
first, reaches a maximum at $V/(Gm_3/a)^{1/2}\simeq2$ and then decreases with increasing
$V$. Note that the impulsive limit, where $\delta e\propto V^{-1}$ (cf.\ Section~3.2 and 
Appendix~A4), would only be reached for much higher values of $V$, and therefore does not
seem astrophysically relevant.

\beginfigure{10}
\epsfxsize 4in
\centerline{
\epsffile{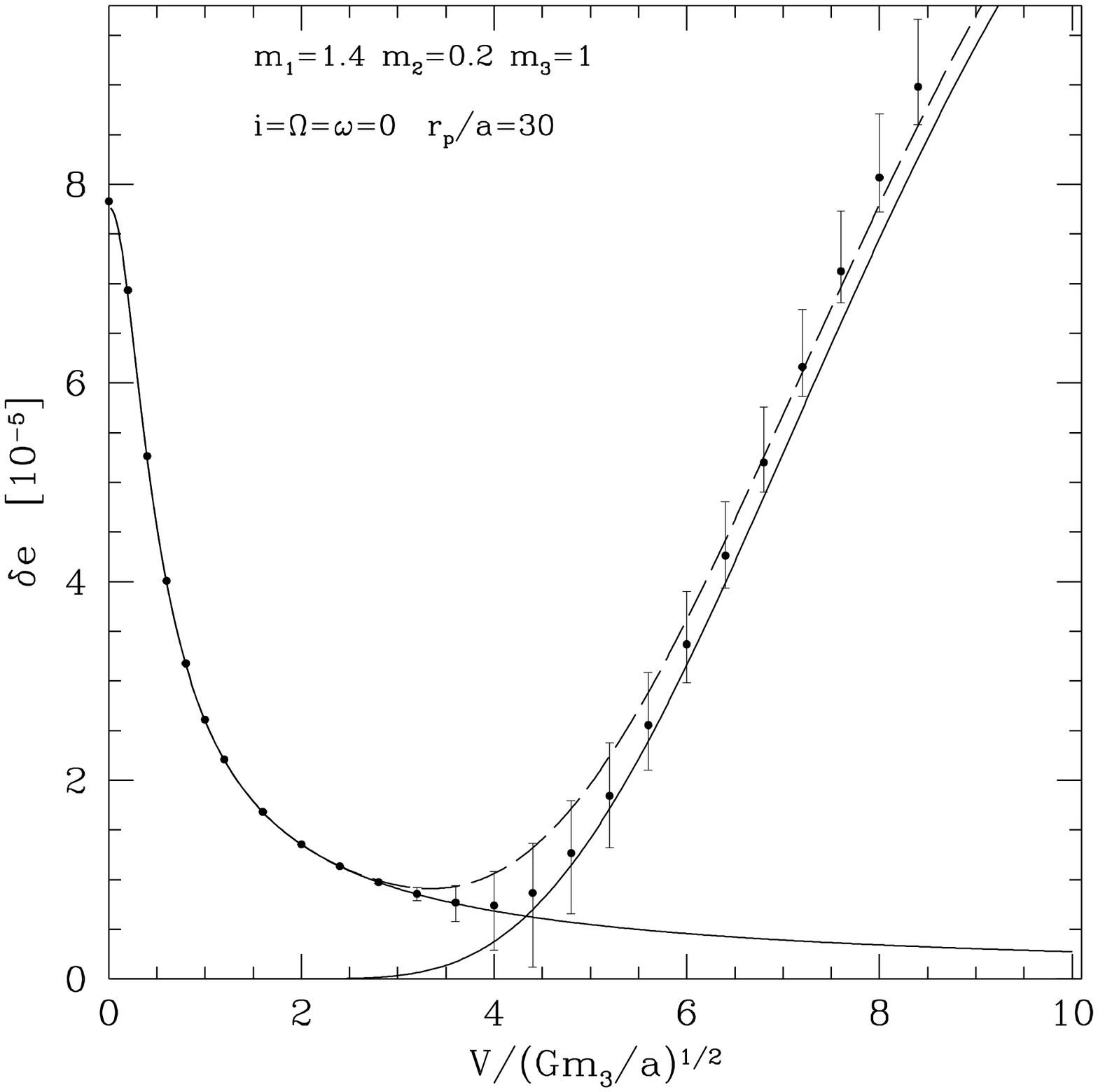}
}
\caption{{\bf Figure 10.} 
Variation of the induced eccentricity $\delta e$ as a function of $V$, the 
relative velocity at infinity for an encounter with fixed $\rp/a=30$.
The two solid lines show the power-law and exponential analytical results
(equations~[9] and~[12]); the dashed line shows their sum. The exponential
result dominates for $V/(Gm_3/a)^{1/2}\ga4$.
}
\endfigure

\beginfigure{11}
\epsfxsize 4in
\centerline{
\epsffile{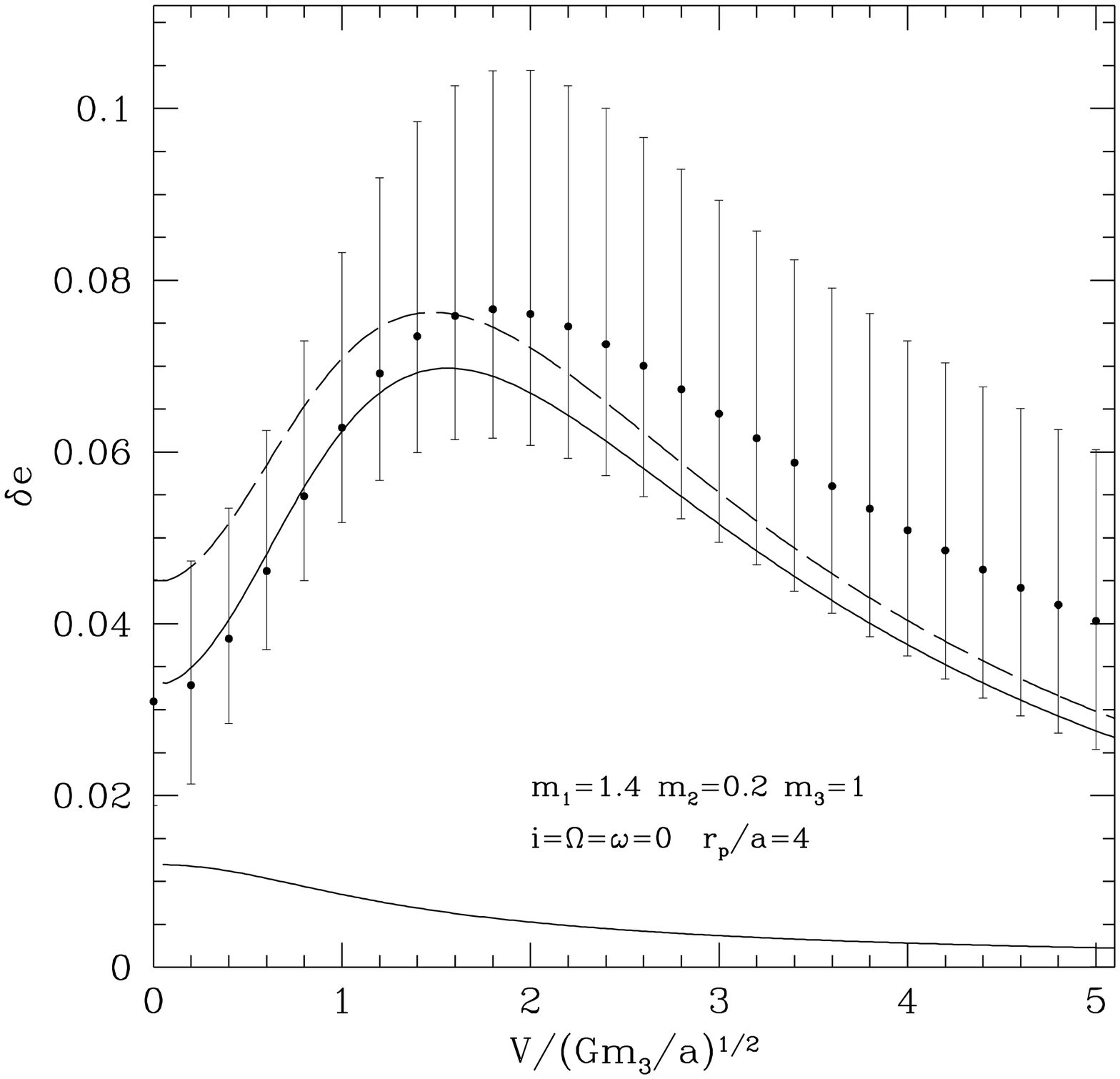}
}
\caption{{\bf Figure 11.} 
Same as Fig.~10 but for $\rp/a=4$. The exponential result (upper solid line) 
dominates everywhere in this case. Notice how the variation of $\delta e$ 
is opposite to
that shown in Fig.~10, with $\delta e$ increasing at first, reaching a maximum
for $V/(Gm_3/a)^{1/2}\simeq2$, then decreasing at higher velocities. 
}
\endfigure

\section{Cross Sections for Eccentricity Change}

\subsection{Analytical Results}

While the results of Section~2 are fully detailed, giving the
dependence on all the circumstances of the encounter, for the
purpose of applications it is necessary to average over several
parameters which are normally only known statistically.  In the
present section we subject the results to some further processing in
order to extract the {\sl cross section\/} for encounters leading to
changes in eccentricity above a given value.  The cross section is
essentially the square of the impact parameter $p$, but
we shall restrict attention to the case of near-parabolic encounters.
In this case the relation between $p$ and the
distance of closest approach reduces to 
$$
\rp\simeq
\displaystyle{{p^2V^2\over 2GM_{123}}}.\eqno\stepeq
$$

\subsubsection{Binaries with Non-Zero Initial Eccentricity}

Consider first binaries with non-zero initial eccentricity.  From
equation~(14) it follows that the dependence of
equation~(8) on the circumstances of the encounter may be summarised as
$$
\delta e = -Ap^{-3}\sin2\Omega\sin^2i,\eqno\stepeq
$$
where 
$$
A = \displaystyle{{15\pi\over4}{m_3M_{123}\over\sqrt{M_{12}}}{(Ga)^{3/2}\over
V^3}e\sqrt{1-e^2}}.\eqno\stepeq
$$  

Now we compute the cross section for events in which $\delta e>\delta
e_0$, where $\delta e_0$ is some value of interest.  
We shall suppose
that $\delta e_0>0$, and then it will be obvious how to extend the
result to negative values, as expression~(8) obviously takes positive and
negative values with equal probability.  Using the appropriate
statistical distributions of $i$ and $\Omega$, we write 
$$
\sigma(\delta e>\delta e_0) = \int 2\pi p dp{1\over2\pi}d\Omega{1\over2}\sin i di,\eqno\stepeq
$$
where the domain of integration is given by $p^3 < -
A\sin2\Omega\sin^2i$, $0<i<\pi$ and, for $\Omega$, the union of the
intervals $[\pi/2,\pi]$ and $[3\pi/2,2\pi]$ (to ensure that
$\delta e$  has the same sign as $\delta e_0$).  The integrations are
straightforward, and lead to the result
$$
\sigma(\delta e>\delta e_0) = \left({A\over\delta
e_0}\right)^{2/3}{9\surd3\over14\pi}
\left[\Gamma\left({2\over3}\right)\Gamma\left({5\over6}\right)\right]^2.\eqno\stepeq
$$
After the definition of $A$ is reinstated the result becomes 
$$
\sigma(\delta e>\delta e_0) =
{9\surd3\over14\pi}\left({15\pi\over4}\right)^{2/3}
\left[\Gamma\left({2\over3}\right)\Gamma\left({5\over6}\right)\right]^2
\left({m_3^2\over M_{12}M_{123}}\right)^{1/3}
{GM_{123}a\over V^2}\,e^{2/3}(1-e^2)^{1/3}\,\delta e_0^{-2/3}.\eqno\stepeq
$$
It is clear that the cross section for negative changes, i.e. $\delta
e<-\delta e_0$, where $\delta e_0>0$, is given by the same expression.

\subsubsection{Initially Circular Binaries}

Now we turn to the cross section corresponding to the parabolic result
for a circular binary in the power-law regime, equation~(11). 
Everything is as straightforward as in
the previous case, with some minor alterations.  First, only positive
values of $\delta e_0$ are relevant.  Second, the appropriate
distribution for $\omega$ is uniform on the range $[0,2\pi]$.  Third,
the integration cannot apparently be performed analytically, and it is
a tricky job numerically because the integrand is not smooth at
certain points in the relevant domain.  For this reason we have opted
for a Monte Carlo evaluation, and so we find that
$$
\sigma(\delta e > \delta e_0) \simeq 4.62 \left({m_3\vert
m_1-m_2\vert\over M_{12}^2}\right)^{2/5}\left({M_{12}\over
M_{123}}\right)^{1/5} {GM_{123}a\over V^2}\,\delta e_0^{-2/5}.\eqno\stepeq
$$
The coefficient is correct to the stated number of figures.

The exponential regime for circular binaries is somewhat trickier,
because the dependence on $\rp$ in equation~(13) is not a simple power law.
Using equation~(14) we first recast equation~(13) as
$$
\delta e = A(Bp^3)^{1/2}\exp(-Bp^3)f_2(i,\omega,\Omega),\eqno\stepeq
$$
where 
$$
A = 3\sqrt{6\pi}m_3/M_{123},~~~~~~~~~~~~~B =
\displaystyle{{1\over3}\left({M_{12}\over
M_{123}}\right)^{1/2}{V^3\over(GM_{123}a)^{3/2}}},\eqno\stepeq
$$
and $f_2$ is the
geometric factor at the end of equation~(13).  Proceeding as in the previous
cases, we can write the desired cross section as 
$$
\sigma(\delta e>\delta e_0) = \int 2\pi p dp {1\over2}\sin i di
{1\over 2\pi}d\theta,\eqno\stepeq
$$
where we have set $\theta = 4\omega + 2\Omega$:  it is clear that the
appropriate distribution is uniform.  The integration with respect to
$p$ gives a factor $\int p dp = p^2/2$, where $p$ is the root of
equation~(21) with $\delta e$ replaced by $\delta e_0$.  It follows that
$$
\sigma(\delta e>\delta e_0) = {1\over4B^{2/3}}\int x^{2/3}\sin i di d\theta,\eqno\stepeq
$$
where $x$ is the larger root of $x - (1/2)\ln x = \ln(Af_2/\delta e_0)$.
An approximate solution is $x = \ln(A/\delta e_0)$, which is valid  when
$\ln(A/\delta e_0)\gg1$, and in that limit we have
$\sigma(\delta e>\delta e_0) = \pi [B^{-1}\ln(A/\delta e_0)]^{2/3}$, i.e.
$$
\sigma(\delta e>\delta e_0) = \pi\left({9M_{123}\over
M_{12}}\right)^{1/3}{GM_{123}a\over
V^2}\left[\ln\left({3\sqrt{6\pi}m_3\over M_{123}\delta
e_0}\right)\right]^{2/3}\chi\left({\delta e_0\over A}\right).\eqno\stepeq
$$
Here $\chi$ is a correction factor which will be used to allow for a
range of values of $\delta e_0/A$; in the limit of $\delta e_0\ll A$ we
use $\chi(0)$ = 1.
For other values of $\delta e_0/A$ we have evaluated the integral in
equation~(24) by a Monte Carlo technique, with results which are given in
Table~1 as the correction $\chi$ to the foregoing asymptotic result.
The data are accurate to 1 unit in the last digit printed.

\begintable{1}
\caption{{\bf Table 1.} Correction to the Asymptotic Cross Section in the
Exponential Regime}
\settabs 11\columns
\+$\displaystyle{{M_{123}\delta e_0\over3 \sqrt{6\pi}m_3}}$
&&$0.333$	&$0.25$	&	$0.2$	&$0.1$ &	$0.033$	&
$0.01$	&	$0.001$	&	$0.0001$	&	$0.00001$\cr
\+$\chi$	&&$0.112$	&$0.228$	&$0.307$	&$0.497$
&$0.705$	&$0.850$	&$0.938$	&$0.968$
&$0.982$\cr
\endtable

The maximum possible value of $\delta e_0$ is $A/\sqrt{2e}$, $\simeq 5.6m_3/M_{123}$.

\subsection{Comparison with Previous Work}

It is interesting to compare these cross sections with those published
by other authors.  

Hut \& Paczy\'nski (1984) gave a result which may
be expressed, using our notation, as
$$
\sigma_{HP}(\delta e>\delta e_0) = (4.7\pm0.9){GM_{123}a\over V^2}
{\sqrt{m_3M_{12}}\over M_{123}} \delta e_0^{-1/3}.\eqno\stepeq
$$
This was obtained by numerical scattering experiments for hard, initially
circular binaries, and so we shall compare it with our equation~(20).  The
result is
$$
{\sigma_{HP}\over\sigma_{HR}} =
(1.02\pm0.19){m_3^{0.1}M_{12}^{1.1}\over\vert
m_1-m_2\vert^{0.4}M_{123}^{0.8}}\delta e_0^{1/15}.\eqno\stepeq
$$
According to Hut \& Paczy\'nski (1984), their result is valid to within the
stated accuracy of 20\% for certain ranges of $a$ and $V$, and for mass ratios in the ranges
$0.03<m_2/m_1<0.3$, $0.2<m_3/m_1<1$.  In this range of masses, the
mass-dependent factor in equation~(27) varies steadily from about $0.59$ (at $m_2 =
0.03 m_1$, $m_3=m_1$) to about $0.95$ (at $m_2=0.3m_1$, $m_3=0.2m_1$).
Actually for the most discrepant result here the scattering cross
section of  Hut \& Paczy\'nski  applies to binaries which include the
borderline between hard and soft pairs, because $m_2$ is so low.
Therefore it is probable that their
cross section is a satisfactory compromise in this domain.  Certainly,
our own result is not intended to be applicable to such encounters.

Now we turn to a detailed comparison with a representative result from
the paper by Rappaport \etal (1989, hereafter RPV). 
They give results for a
single-parameter family of circular binaries in each of two
environments, characterised by mixtures of stars with given mass,
number density and velocity dispersion.  We have chosen to compare
with their result for the hardest binary (one of orbital period 3
days) in their model for the stellar population of the star cluster
$\omega$ Cen.  Our first step is to average our result, equation~(20), over
a Maxwellian distribution of velocities, though with an extra factor
of $V$ in the averaging to account for the enhanced rate of
interaction between stars of high relative speed (as was done by RPV). 
The result is that the factor $V^2$ in equation~(20) is replaced by
$2(\langle v_b^2\rangle + \langle v_3^2\rangle)/3$, where the averages
are the mean square three-dimensional speeds of the binary and the
third body, respectively.  Next the cross section is averaged over
the stellar species in the mixture specified by RPV.  Then
a similar process is carried out for the cross section, equation~(25),
corresponding to the exponential regime.  This is slightly more
awkward, as the averaging over the stellar species must be carried out
afresh for each value of $\delta e_0$. For comparison with the results of RPV
we simply adopt the larger of our two cross sections at each
value of $\delta e_0$.

Fig.~12 shows the ratio of RPV's cross section to ours as a function of $\delta e_0$.
The change in slope near $\delta e_0\simeq 0.003$ is associated with the
point at which the cross section of equation~(25) begins to dominate that
of equation~(20).  The change at $\delta e_0= 0.01$ 
has a similar origin in the work of RPV:  the functional
form of their cross section changes at this point.  From there up to
the point where $\delta e_0\simeq0.5$ our cross section is
systematically higher, by  about 60\%, than that of RPV.
The reason for this systematic disagreement is not certain, but it is
clear that our approach of using the maximum cross section at each
value of $\delta e_0$ is rather crude.
The sharp drop towards $0$ as  $\delta e_0\to1$ reflects the fact
that our cross section does not vanish in this limit, whereas that of
RPV does so. Turning now to very small values of $\delta e_0$, we see that
the cross section of RPV decreases monotonically relative
to ours. Indeed, their expressions are similar to our equation~(25),
valid in the exponential regime but not in the power-law regime. 
This may be related to their use of relatively short and low-precision 
numerical integrations,
which are probably inadequate for computing the very small changes in eccentricity
corresponding to the power-law regime. It must also be pointed
out, however, that our cross sections were calculated for parabolic encounters,
which is inappropriate when $\rp/a\ga
GM_{123}/(aV^2)$. Substitution of typical values into equation~(11) therefore
shows that the validity of our
result is restricted to the range $\delta e_0\ga 10^{-5}$ in this case.

\beginfigure{12}
\epsfxsize 4in
\centerline{
\epsffile{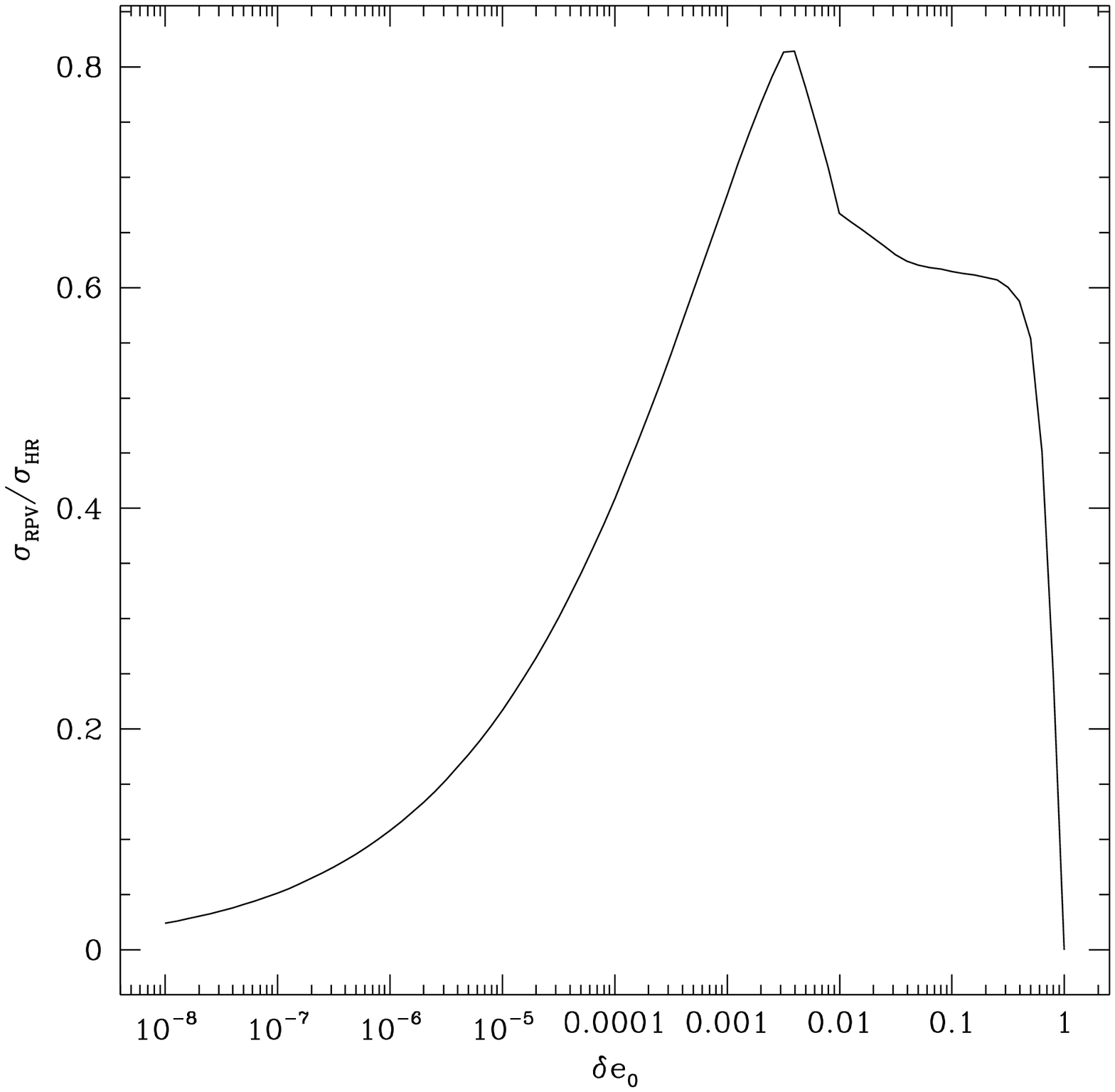}
}
\caption{{\bf Figure 12.} 
Comparison between the scattering cross sections $\sigma(\delta
e>\de_0)$.  The result of Rappaport \etal (1989), denoted by the
subscript RPV, is compared with the cross section derived from the
results of this paper, for a range of values of $\de_0$.}
\endfigure

Next we turn to the work of Hills (1991), who computed cross
sections for equal-mass binaries in encounters with more massive
intruders ($m_3\gg m_1=m_2$).  The initial eccentricity was
$e=10^{-4}$, and since Hills' results correspond to much larger
changes, in the range $\delta e>0.1$, we shall compare equation~(25) 
with his results extrapolated to $V=0$.  He gives a scaled
cross section $\sigma(\delta e>0.3)$ for mass ratios in the range
$10<m_3/m_1 <10^4$. Our results show the same trend with
increasing mass ratio, but are consistently smaller by about
10--20\%.  The dependence of the cross section on $\delta e_0$
can also be checked, as Hills gives the ratios of $\sigma(\delta
e>\delta e_0)$ for the values $\delta e_0 = 0.1$, $0.3$ and
$0.7$.  For the smaller values of $\delta e_0$ our result agrees
with Hills' to within the same accuracy of 10--20\%.   For the
larger values the agreement is poorer: Hills finds values of
$\sigma(\delta e>0.3)/\sigma(\delta e>0.7)$ in the range
$1.5$--$3$, whereas our results yield a ratio closer to $1.4$.
Since our results are based on assumptions which are valid for
small eccentricity changes, it is perhaps surprising that the
agreement is so good.

Finally we compare our results with those of Phinney (1992), who
gives formulae for the induced eccentricity in an initially
circular binary, and an approximate cross section.  
Like us, Phinney assumes that $\rp\gg a$. For simplicity, he only
considers the coplanar case and he assumes a straight-line (i.e.,
extremely hyperbolic) trajectory. In this limiting case, the exponential
result always dominates and the appropriate
comparison is to our equation~(12) in the limit where 
$e^\prime\rightarrow\infty$ and for $i=0$. Identifying Phinney's quantity
$\lambda$ with $(M_{12}/M_{123})^{1/2}(r_p/a)^{3/2}(e^\prime)^{-1/2}$
in our notations, we find perfect agreement with Phinney's equation~(5.2).
Note that Phinney's equations~(5.2) and~(5.3) have been mislabelled: they
correspond to the prograde and retrograde cases, respectively (Phinney,
personal communication). Phinney's result for the coplanar retrograde case 
is of higher order and was neglected in our derivation of equation~(12).

Phinney (1992) also gives a result for fast encounters (the ``impulsive limit''), 
which we now consider briefly, although we do not expect this case to be of
importance in most astrophysical applications (one possible exception being
the perturbation of wide circular binaries by halo stars in the Galaxy).   
We give a derivation in Appendix A4, and our
result for coplanar encounters with an initially circular binary is
$$
\delta e = {2 G^{1/2} m_3 a^{3/2}\over
M_{12}^{1/2}V\rp^2}\vert\cos\theta\vert\left(\cos^2\theta +
4\sin^2\theta\right)^{1/2},\eqno\stepeq
$$
where $\theta$ is the angle between the vectors $\br$ and $\bR$ at the
time of pericentre passage. 
This result vanishes when $\theta = 90^\circ$, which can be
understood because in that case the third body generates equal impulses
in both components, and therefore does not alter their relative motion. 
Phinney's equation~(5.1) as published differs in the trigonometrical factor.
This is because of a typographical error: the first term inside the
square root, which is printed as $3\sin^2  \theta_0$ should have been
$3\sin^2 2\theta_0$ (Phinney, personal communication).
With this typo corrected, it is easy to show that Phinney's equation~(5.1) is 
in fact identical to our equation~(28).

\section{Conclusions and Discussion}

We now summarise our main results.  They describe the change $\delta e$ in the
eccentricity of a binary of semi-major axis $a$ which experiences
an encounter with a third body, in the following regime:  if $\rp$ is
the distance of closest approach of the perturber, we suppose that
$\rp$ considerably exceeds $a$ and that the time scale of the encounter
considerably exceeds the period of the binary.  With these assumptions
our results are of most relevance to hard binaries, and the orbit of the
third body is nearly parabolic.  

\beginlist

\item{1.} When the induced change in eccentricity, $\delta e$, is small
compared with the initial eccentricity of the binary, then
$\delta e$ falls off as $(a/\rp)^{3/2}$.  The relevant results
are equation~(7) for a hyperbolic encounter and equation~(8) in
the parabolic limit.  Other factors affecting the result are the
masses of the participants, the geometry of the encounter, the
initial eccentricity of the binary and the orbit of the
perturber.  The result comes from the quadrupole interaction
between the binary and the third body.  Its main relevance is to
binaries with a large initial eccentricity (Fig.~1).

\item{2.} When  $\delta e$ is large compared with
the initial eccentricity, and $\rp$ much exceeds $a$, the final
eccentricity falls off in general as $(a/\rp)^{5/2}$ (equations~[9] and~[11]), and also
depends on various other factors.  The fact that this result originates
in the octupole interaction accounts for the extra power.  This result
is of relevance mainly to binaries whose initial eccentricity is small (e.g.,
initially circular binaries), except for very close
encounters (cf.\ Figs.~2 and~4).

\item{3.} When $\rp$ exceeds $a$, but not by a large factor, the induced
eccentricity falls off exponentially with a power of $\rp/a$,
though the coefficient also depends on this ratio, and other
factors (equations~[12] and~[13]).  The exponential factor occurs
because, at these values of $\rp/a$, the time scale of the
encounter begins to be comparable with the period of the binary.
It is mainly of relevance to binaries with small initial
eccentricity, and to encounters which lead to a large induced
eccentricity, but at periastron distances $\rp$ where the
quadrupole interaction is still dominant.  It is also relevant to
the very special case of a binary with components of equal mass
(Fig.~3), or to the more important case of a massive perturber
(Fig.9). 

\item{4.} Cross sections corresponding to these three results are given
in equations~(19), (20) and~(25).  The first two depend on certain powers of
the change in eccentricity, while the dependence in the third is
approximately logarithmic. 

\endlist

These results have been compared with numerical data in various ways. 
We have tested the predicted dependence of the induced eccentricity
change $\delta e$ on the encounter distance
$\rp$ (Figs.~3, 4 and~7), on the masses of the stars (Figs.~8 and~9), 
and on the geometry of the encounter (Figs.~5, 6, 10
and~11). We have almost always found excellent agreement between numerical
and analytical results.

We have also compared our results, and
especially the cross sections, with those derived by other authors.  
The main point at which previous work requires revision is the behaviour
of initially circular binaries.  Previously it was thought that the
induced eccentricity $\delta e$ falls off exponentially with encounter distance,
but it is now clear that this result applies only to a limited range of
periastron distance, beyond which $\delta e$ falls off only as a power of the
distance. The effect of this is that the cross section for encounters
leading to small (but observable) induced eccentricities is very
considerably larger than was previously sometimes thought.

The advantage of using our formulae for the induced eccentricity change
in future investigations is speed.  Where the parabolic approximation is
valid, the stated cross sections may be used directly, and it is a
straightforward matter to compute the cross section for any reasonable
mass spectrum.  Where the parabolic approximation is invalid, but the
encounters are still both slow and tidal, cross sections may be computed
in a Monte Carlo fashion, the outcome of each encounter being obtained
from the appropriate formulae of Section~2.  In this way estimates of cross
sections in a wide variety of collisional stellar systems may be
estimated much more rapidly than in the past, without the need for
numerous careful and time-consuming three-body integrations.

Currently the most important application of our results concerns low-mass
binary millisecond pulsars (LMBPs) in globular clusters.
These are formed when an old neutron star in a binary system 
accretes material from a red-giant companion. 
This leads to the production of a spun-up (recycled) pulsar with a low-mass
white-dwarf companion in a wide, circular orbit (typical eccentricities at
birth being in the range $e_i\sim10^{-6}-10^{-3}$).
This formation process is well understood theoretically 
(Verbunt 1993; Phinney \& Kulkarni 1994).
In globular clusters, the wide orbits of LMBPs can be perturbed significantly 
by passing stars. Thus final eccentricities $e_f \gg e_i$ can be observed today,
and the measured values contain important information about the dynamical history
of the binaries and their environment. In fact
Rasio \& Heggie (1995) have shown, using the analytical results derived here,
that {\sl all\/} currently known LMBPs
in globular clusters must have been affected by
interactions, with their current eccentricities being at least an order of magnitude
larger than at birth.

Distant interactions with passing stars may also be important for
the orbital evolution of X-ray binaries in globular clusters.
Hut \& Paczy\'nski (1984) have shown that even a very small
change in eccentricity can lead to a catastrophic increase in the
mass transfer rate in a semidetached binary. In some cases, an
eccentricity change as small as $\delta e\sim10^{-4}-10^{-3}$
could even lead to the formation of a common envelope, ultimately
destroying the binary.  Other problems where our results could
find applications include the interactions of binaries with
massive black holes (Hills 1991) and the orbital evolution of a
massive black-hole binary in a galactic nucleus (Mikkola \&
Valtonen 1992). These problems involve extreme mass ratios
($m_3\gg M_{12}$ or $m_3\ll M_{12}$) which we did not address
specifically in this paper, but our analytical expressions should
apply in those cases as well (cf.\ Fig.~9).

\section*{Acknowledgements}

We are much indebted to Steve McMillan, whose numerical work put our
investigations on the right track. We thank Sterl Phinney for useful
comments on the manuscript and Piet Hut and Scott Tremaine for 
helpful discussions. This work was supported in part by  
NASA Grant HF-1037.01-92A while F.A.R.\ was a Hubble Fellow at the Institute
for Advanced Study, Princeton.
D.C.H.\ thanks the Institute for Advanced Study for its hospitality 
while work on this project was being carried out.

\appendix

\section{Derivation of the Analytical Results}

\subsection{Nonzero Initial Eccentricity}

Our task here is to derive equation~(7) by integration of equation~(6), where
$\bF$ is given by equation~(4).  At
lowest order, which is all we shall need,  we
substitute unperturbed expressions for $\br$ and $\bR$ on the right
side of equation~(4). 

We denote by $\bahat$ and $\bbhat$ two orthogonal unit vectors in
the original plane of motion of the binary, such that $\bahat$ is in the
direction of pericentre.  (In fact $\be = e\bahat$.)  We also let $\bchat =
\bahat\times\bbhat$ be the unit vector in the direction of the angular
momentum. 
Similarly
we define two vectors $\bAhat$ and $\bBhat$ in the original plane of
motion of $m_3$ relative to the barycentre of the binary.  

Next, define
coordinates $x, y, X$ and $Y$ such that $\br = x\bahat + y\bbhat$ and 
$\bR = X\bAhat + Y\bBhat$.   Then expressions for these four coordinates
in unperturbed Keplerian motion can be found in standard texts (e.g.,
Plummer 1918; Danby 1962).  Thus $x = a(\cos E - e)$ and $y = b\sin E$
where $b = a\sqrt{1-e^2}$ and $E$ is the eccentric anomaly.  It is
related to time by Kepler's equation, 
$n(t-t_0) = E - e\sin E$, where
$$
n^2a^3 = GM_{12},\eqno\stepeq
$$ 
and $t_0$ is the time of pericentre passage. If the relative motion of the 
third body is hyperbolic with eccentricity $e^\prime$, for this section and 
the next one the  expressions are most
conveniently written in polar form, i.e., $X = R\cos\theta$ and $Y =
R\sin\theta$.   Here the polar angle is related to time implicitly via
angular momentum conservation, i.e. $R^2\dot\theta =
[GM_{123}\rp(e^\prime + 1)]^{1/2}$,  where $\rp$ is the distance of
closest approach of the third body to the barycentre of the binary.  We
take $t = 0$ at this instant, i.e., when $\theta = 0$.  Also $R =
\rp (e^\prime +1)/(e^\prime\cos\theta +1)$.

The integration of the equation for $\dot\be$ proceeds by first
averaging over the motion of the binary (at fixed $\bR$). At this point,
if only the quadrupole term ($n = 2$) is retained in equation~(4),
the result is
$$
\langle\dot\be\rangle = -{\pi m_3\over M_{12}R^5}{a^2be\over T}
\left[ 15\xi\eta\,\bahat +
(-9\xi^2 + 6\eta^2 +3\zeta^2)\,\bbhat + 3\eta\zeta\,\bchat\right],\eqno\stepeq
$$
where $T$ is the period of the binary, and $\xi, \eta$ and $\zeta$ are
defined implicitly by $\bR = \xi\bahat + \eta\bbhat +
\zeta\bchat$.  In fact only the component of equation~(6) along
$\bahat$ is needed, since $e\delta e = \be\cdot\delta\be = e\bahat\cdot\delta\be$;
and since $Y$ is an odd function of $t$
we find that the change in eccentricity is
$$
\delta e = - {15\pi m_3\over M_{12}}{a^2be\over T}\left(\bahat\cdot\bAhat\, \bbhat\cdot\bAhat
\int_{-\infty}^\infty {X^2\over R^5}dt + \bahat\cdot\bBhat\, \bbhat\cdot\bBhat 
\int_{-\infty}^\infty {Y^2\over R^5}dt\right).\eqno\stepeq
$$

For hyperbolic motion of the third body the result is 
$$\eqalign{
\delta e = &-{5e\over 2e^{\prime^2}} \sqrt{{m_3^2a^3(1-e^2)\over
M_{12}M_{123}\rp^3(1+e^\prime)^3}}\times\cr
&\times
\left\{ \bahat\cdot\bAhat\, \bbhat\cdot\bAhat \left[3e^{\prime^2}\arccos\left(-{1\over
e^\prime}\right)  + (4e^{\prime^2}-1)\sqrt{e^{\prime^2}-1}\right] 
 + \bahat\cdot\bBhat\, \bbhat\cdot\bBhat \left[3e^{\prime^2}\arccos\left(-{1\over
e^\prime}\right) + (2e^{\prime^2} +
1)\sqrt{e^{\prime^2}-1}\right]\right\}\cr
}\eqno\stepeq
$$

For parabolic motion ($e^\prime\to1$) the corresponding result is 
$$
\delta e = -{15\pi\over 8} \left({2 m_3^2 a^3\over M_{123} M_{12}
\rp^3}\right)^{1/2}e\sqrt{1-e^2}\,
(\bahat\cdot\bAhat\, \bbhat\cdot\bAhat + \bahat\cdot\bBhat\, 
\bbhat\cdot\bBhat),\eqno\stepeq
$$
a result already given in Heggie (1975, eq.~[5.66]), though with the wrong
sign.  It is convenient to write the direction cosines $\bahat\cdot\bAhat$, 
$\bbhat\cdot\bAhat$, etc., in terms of the Keplerian elements outlined in
the text. This leads to equations~(7) and~(8).

\subsection{Initially Circular Binaries: The Power-Law Regime}

Both of these results vanish for an initially circular orbit.  However, the
octupole ($n = 3$) terms in equation~(4) yield a non-trivial result in this
case, which is the case we consider henceforth.  As before we take $t=0$
at the instant of closest approach of $m_3$, but now we can choose
$\bahat$ to be in the direction of $\br$ at that time.  The analogue of
equation~(A2) is 
$$
\langle\dot\be\rangle = {15m_3(m_1-m_2)\pi a^4\over
8M_{12}^2TR^5}\left(1-5{\zeta^2\over R^2}\right)(\eta\bahat - \xi\bbhat).
$$
Converting to coordinates in the plane of motion of the third body, we
find that
$$
\delta\be =  {15m_3(m_1-m_2)\pi a^4\over 8M_{12}^2T}
\bigg(\int_{-\infty}^\infty{X^3\over
R^7}dt\, [1-5(\bAhat\cdot\bchat)^2]\,\bAhat\times\bchat +
\int_{-\infty}^\infty{XY^2\over
R^7}dt\,\left\{[1-5(\bBhat\cdot\bchat)^2]\,\bAhat\times\bchat -
10\bBhat\cdot\bchat\, \bAhat\cdot\bchat\, 
\bBhat\times\bchat\right\}\bigg).\eqno\stepeq
$$
Note that $\delta\be$ lies in the plane of motion of the binary, and is
independent of the choice of $\bahat$ and $\bbhat$ in this plane.
Incidentally, it is also easily seen that $\delta\be$ is orthogonal to
$\bAhat$ in certain circumstances, e.g., in coplanar motion.  This means
that the binary orbit becomes elongated in a direction orthogonal to
the position of $m_3$ at pericentre.

For hyperbolic relative motion of the third body the two integrals in
equation~(A6) have the values
$$
\int_{-\infty}^\infty {X^3\over R^7} dt = {\sqrt{e^{\prime^2}-1}
(2+11e^{\prime^2}+32e^{\prime^4})+45e^{\prime^4}\arccos(-1/e^\prime)\over
30e^{\prime^3}\sqrt{GM_{123}\rp^5(e^\prime+1)^5}}
$$
and
$$
\int_{\infty}^\infty {XY^2\over R^7} dt = {\sqrt{e^{\prime^2}-1}
(-2+9e^{\prime^2}+8e^{\prime^4})+15e^{\prime^4}\arccos(-1/e^\prime)\over
30e^{\prime^3}\sqrt{GM_{123}\rp^5(e^\prime+1)^5}}.
$$

  From these results the values of $\delta\be$ and $\delta e = 
\vert\delta\be\vert$ are easily derived, though the expressions are
relatively cumbersome.  It helps to resolve $\delta\be$ along and
perpendicular to the line of nodes, and to express the various products
of unit vectors in terms of the elements $i$ and $\omega$ already 
introduced in \S2.2.
Eventually we obtain equation~(9), where $f_1(e^\prime)$ is defined in 
equation~(10).
Note that, in the case of parabolic motion of the third body, we have
$f_1(1) = \pi$, and so the result is rather more
graceful.

\subsection{Initially Circular Binaries: The Exponential Regime}

As we have seen, averaging over the fast motion of the
binary leads to the conclusion that the quadrupole term gives a null
result in the case of an initially circular binary, 
and that the octupole term provides the
leading term in the induced eccentricity. At a sufficiently small
distance of closest approach, however, the method of averaging becomes
inaccurate, and so a different approach to the evaluation of the
quadrupole term is needed.  We continue, however, to restrict attention
to an initially circular orbit.

In this case the contribution from the quadrupole term is  
$$ 
\dot\be =
{m_3\over M_{12}R^3}\left[6{(\br\cdot\bR)(\dot\br\cdot\bR)\over R^2}\br -
3{(\br\cdot\bR)^2\over R^2}\dot\br + r^2\dot\br\right].\eqno\stepeq
$$ 
As before we take
$t=0$ at the time of closest approach of $m_3$, and choose the vector
$\bahat$ to be the direction of $\br$ at this instant, so that $\br = 
a(\bahat\cos nt + \bbhat\sin nt)$, where $n$ is the mean motion of the
binary.  

Since the right side of equation~(A7) involves the third power of
components of $\br$ or $\dot\br$, its terms involve factors of $\exp(\pm int)$ and
$\exp(\pm3int)$  when written in exponential form.  From what follows we
shall see that the resulting expressions for $\delta e$ would decrease like
$\exp(-A)$ and $\exp(-3A)$, respectively, where $A$ is proportional to
$(\rp/a)^{3/2}\gg1$.  Therefore we shall ignore at the outset the terms
in $\exp(\pm3int)$, which simply means that the method of averaging
applies to them with greater accuracy because of their faster frequency.
At this stage, then, we have
$$
\dot\be = {m_3a^3n\over M_{12}R^5}\,\Re\left[ e^{int}\bahat(iR^2 + {3\over4}i\xi^2
+ {9\over2}\xi\eta - {15\over4}i\eta^2) 
+ e^{int}\bbhat(R^2 - {15\over4}\xi^2 + {9\over2}i\xi\eta + {3\over4}\eta^2) \right];
\eqno\stepeq
$$
here, as before, we have written $\bR = \xi\bahat + \eta\bbhat +
\zeta\bchat$, where $\bchat = \bahat\times\bbhat$.

When we integrate with respect to $t$ we require integrals of the form 
$$
I_1 = \int_{-\infty}^\infty {X_iX_j\over R^5}\exp(int)\,dt,\eqno\stepeq
$$
where $X_i$ and $X_j$ are two of the non-zero components of $\bR =
X\bAhat + Y\bBhat$.  Using two standard equations for hyperbolic motion, viz.
$$
n^\prime t = e^\prime\sinh F - F,\eqno\stepeq
$$ 
and 
$$
R = a^\prime(e^\prime\cosh F - 1),\eqno\stepeq
$$ 
where 
$$
n^{\prime^2}a^{\prime^3} = GM_{123},\eqno\stepeq
$$
we write this as 
$$
I_1 = {1\over n^\prime a^\prime}\int_{-\infty}^\infty 
{X_iX_j\over R^4}\exp\left[ i {n\over n^\prime}(e^\prime\sinh F - F)\right]\,dF.
\eqno\stepeq
$$

Now we recall the assumption that the encounter is slow (eq.~[2]), which
implies that the 
exponent here is rapidly oscillating compared with the time scale of
variations of the other factors in the integrand.  The asymptotic
treatment of such integrals is standard.
The idea is to evaluate the integral approximately by deforming the
contour of integration off the real axis to the saddle point of the
exponent (``method of steepest descents''), which occurs in this case 
at $R = 0$, by
equation~(A11).  (The process is rather like that adopted by Heggie 1975 in
evaluating the change in energy, and the relevant contour is illustrated
there.) The rest of the integrand in equation~(A13) is singular here,
however, and so we first reduce the singularity using partial
integration.  From equation~(A11) we have 
$\displaystyle{{dR\over dF} = a^\prime
e^\prime\sinh F}$, and so equation~(A13) becomes
$$
I_1 = {1\over 3n^\prime {a^\prime}^2 e^\prime}\int_{-\infty}^\infty {1\over
R^3}
{d\over dF}\left\{ {X_iX_j\over\sinh F}\exp\left[i{n\over
n^\prime}(e^\prime\sinh F - F)\right]\right\}\,dF,
\eqno\stepeq
$$
though we have already had to deform the contour to avoid the
singularity at $F = 0 $.  In performing the differentiation, to leading order
we need only
include the derivative of the exponential, since the
factor $n/n^\prime $ brings down a factor  of order
$(\rp/a)^{3/2}$.  To leading order we therefore have
$$
I_1 = {in\over 3GM_{123}e^\prime}\int_{-\infty}^\infty{X_iX_j\over R^2\sinh F}
\exp\left[ i {n\over n^\prime}(e^\prime\sinh F - F)\right]\,dF,
\eqno\stepeq
$$
where we have made use of equations~(A11) and~(A12).

A second application of the same trick leads to 
$$
I_1 = -{n^2\over 3GM_{123}a^{\prime^2}n^\prime e^{\prime^2}}
\int_{-\infty}^\infty {X_iX_j\over \sinh^2F}
\exp\left[ i {n\over n^\prime}(e^\prime\sinh F - F)\right]\,dF,
\eqno\stepeq
$$
and now we turn attention to the saddle point at $R = 0$, i.e., 
$F = i\arccos(1/e^\prime)$. In its vicinity we have 
$$
e^\prime\sinh F - F \simeq i\sqrt{e^{\prime^2}-1} -
i\arccos{1\over e^\prime} + {i\over 2}(F - i\arccos{1\over
e^\prime})^2\sqrt{e^{\prime^2}-1}.\eqno\stepeq
$$
Now the integration is trivial, since in the rest of the integrand we
can substitute values at the saddle, and so 
$$
I_1 = {\sqrt{2\pi}\over3}{M_{12}^{3/4}\over G^{1/2}M_{123}^{5/4}}
{X_iX_j\over a^{9/4}\rp^{5/4}(e^\prime+1)^{5/4}}
\exp\left[ -\left({M_{12}\over
M_{123}}\right)^{1/2}\left({\rp\over a}\right)^{3/2}{\sqrt{e^{\prime^2}
-1}-\arccos(1/e^\prime)\over(e^\prime-1)^{3/2}}\right],\eqno\stepeq
$$
though we have yet to substitute values for $X_i$ and $X_j$ at the saddle
point, where
$$
\bR = {\rp(e^\prime+1)\over e^\prime}(\bAhat + i\bBhat).\eqno\stepeq
$$

Now we return to equation~(A8).   
Using equations~(A18) and~(A19) we readily find that 
$$\eqalign{
\be = &{\sqrt{2\pi}m_3 M_{12}^{1/4}\rp^{3/4}(e^\prime +1)^{3/4}\over
3M_{123}^{5/4}a^{3/4}e^{\prime^2}}\times
\exp\left[ -\left({M_{12}\over
M_{123}}\right)^{1/2}\left({\rp\over a}\right)^{3/2}{\sqrt{e^{\prime^2}
-1}-\arccos(1/e^\prime)\over(e^\prime-1)^{3/2}}\right]\times\cr
&\times\bigg\{\bahat\left[ -{3\over2}\bahat\cdot\bAhat\, \bahat\cdot\bBhat +
{9\over2}(\bahat\cdot\bAhat\, \bbhat\cdot\bAhat - \bahat\cdot\bBhat \bbhat\cdot\bBhat) +
{15\over2}\bbhat\cdot\bAhat\, \bbhat\cdot\bBhat\right] +\cr
&~~~~ + \bbhat\left[ -{15\over4}(\bahat\cdot\bAhat)^2 + {15\over4}(\bahat\cdot\bBhat)^2 -
{9\over2}(\bahat\cdot\bAhat\, \bbhat\cdot\bBhat + \bahat\cdot\bBhat\, \bbhat\cdot\bAhat) +
{3\over4}(\bbhat\cdot\bAhat)^2 - {3\over4}(\bbhat\cdot\bBhat)^2\right]\bigg\}.\cr
}\eqno\stepeq
$$

The scalar products here are easily evaluated in terms of the usual
Keplerian elements $\omega,\ \Omega$ and $i$, where $\Omega$ is measured
along the plane of motion of the binary from the direction of $\bahat$
to the ascending node, and $\omega$ is measured from the node to the
direction of $\bAhat$.  The last (geometric) factor in expression~(A20) reduces,
after some lengthy expressions, to the surprisingly compact expression
$$
\bahat\left[ -{3\over2}\sin2\omega\sin^2i +
{9\over4}\sin2(\Omega+\omega)\,(1+\cos i)^2\right]
+\bbhat\left[ -{3\over2}\cos2\omega\sin^2i -
{9\over4}\cos2(\Omega+\omega)\,(1+\cos i)^2\right],\eqno\stepeq
$$
whence the magnitude of the vector $\be$ is given by equation~(12).
When $e^\prime = 1$ (parabolic motion) the exponential factor is
$\displaystyle{-{2\over3}\left({2M_{12}\over
M_{123}}\right)^{1/2}\left({\rp\over a}\right)^{3/2},}$ and so we have
arrived at equation~(13).

\subsection{Initially Circular Binaries: The Impulsive Limit}

We consider briefly the case in which the inequality in equation~(2) is
reversed.  The duration of the encounter is then very short compared
with the binary period.  This case appears unimportant for most astrophysical
applications, and we include it here mainly for the discussion in Section~3.2.

In equation~(6) we model $\bF$ as a delta function, and then this equation
integrates to 
$$
GM_{12}\be = 2(\delta\dot\br.\dot\br)\br -
(\delta\dot\br.\br)\dot\br,\eqno\stepeq
$$ 
where $\delta\dot\br$ is the change in
relative velocity of the components due to the encounter.  (The second
term in eq.~[6] vanishes for a circular binary.)  We evaluate $\dv$ by
integrating the quadrupole term of equation~(4), keeping $\br$ fixed and
taking for $\bR$ the form corresponding to a rectilinear orbit.  In
the notation of Appendix~A1 this is
$\bR = \rp\bAhat + Vt\bBhat$, if $t=0$ at the time of pericentre.  
It follows that 
$$
\dv_i = {Gm_3r_j\over V\rp^2}(4\hat A_i\hat A_j + 2\hat B_i\hat B_j - 
2\delta_{ij}),\eqno\stepeq
$$
where summation over $j$ is implied, and the $\delta$ on the right 
is the Kronecker symbol.  

When this result is substituted
into equation~(A22) we also write $\br = a\bahat$ and 
$\dot\br = na\bbhat$, where
the unit vectors have the meaning assigned to them in \S A2, 
and $n$ is
defined in equation~(A1).  It quickly
follows that 
$$
GM_{12}\be = {Gm_3na^3\over V\rp^2}\left\{ 4 (2\bahat\cdot\bAhat\, \bbhat\cdot\bAhat +
\bahat\cdot\bBhat\, \bbhat\cdot\bBhat) \bahat - 2\left[ 2(\bahat\cdot\bAhat)^2 +
(\bahat\cdot\bBhat)^2 - 1\right]\bbhat\right\}.\eqno\stepeq
$$
The two terms are orthogonal, and so $e$ is easily found.  

In the case of coplanar motion, we introduce the angle $\theta$, defined
to be the angle between the vectors $\bahat$ and $\bAhat$. Then the
direction cosines are easily expressed in terms of $\theta$, and equation~(28)
follows easily.

\section{The Equal-Mass Case}

The purpose of this appendix is to show that, for a circular binary with
components of equal mass $m$, the induced eccentricity vanishes to all orders of the
perturbation parameter $a/\rp$.  In first order perturbation theory this
is easy to understand.  All terms in equation~(4) with $n$ odd contribute
nothing if $m_1=m_2$.  Hence $\bF$ contains only odd monomials in the
components of the vector $\br$, and then it is easy to see that the
right side of equation~(6) vanishes after averaging over one binary period, if
$e=0$. The
following argument  extends this reasoning to perturbation theory at all
orders.

Higher-order perturbation theory is best addressed using Hamiltonian
methods, and here we describe the appropriate variables in the simplest
non-trivial case: coplanar motion of all three stars.  The relative
motion of the binary components is conveniently handled in
action-angle variables, which, for planar Keplerian motion, can be taken
as equivalent to the classical Delaunay variables $L,~G,~l$ and $g$ 
(cf.\ Plummer 1918). 
The variables $L$ and $G$ are expressible in terms of the semi-major
axis $a$ and eccentricity $e$ by means of 
$$
a = 2L^2/(Gm^3)\eqno\stepeq
$$ 
and 
$$
G =
L\sqrt{1-e^2}.\eqno\stepeq
$$  
The conjugate variable $l$ is the mean anomaly, which
increases uniformly with time in unperturbed motion, passes the value $0$ at pericentre, and
increases by $2\pi$ in each revolution. The variable $g$ is the
longitude of the pericentre, which we assume is measured from the
direction of one axis of an inertial frame.  The Hamiltonian function
for unperturbed Keplerian motion is a function of $L$ alone, denoted by
$H_b(L)$.

The motion of the third star we continue to describe by its position
vector $\bR$ with respect to the barycentre of the binary, and we
introduce the conjugate momentum $\bP$.  Then the Keplerian
approximation to its motion can be described by a Hamiltonian function
$H_3(\bR,\bP)$.

The whole Hamiltonian can now be written as 
$$
H = H_b(L) +
H_3(\bR,\bP) + \Re(\br(L,G,l,g),\bR),\eqno\stepeq
$$
where the three
terms on the right correspond, respectively, to the binary,
the Keplerian motion of the third body, and the perturbation.
The last of these terms is, except for the terms with $n = 0$ and $1$
and a factor involving the masses,
nothing other than the expression whose gradient appears in  
equation~(4), but we write $\br$ in terms of the Delaunay variables.

In the case of equal masses, $\Re$ is invariant under the transformation
$\br\rightarrow-\br$, i.e. the transformation $g\rightarrow g+\pi$, with the
other three canonical variables unchanged.  To see that these
transformations are equivalent, it is useful to consider the case of a
binary of low eccentricity.  Approximate expressions for the relative 
position vector $\br$ of the two
components are easily found (cf.\ Plummer 1918 again) as
$$
\br = a(\cos l - {3\over2}e + {1\over 2}e \cos 2l, \sin l +
{1\over2}e\sin 2l),\eqno\stepeq
$$
if the components are resolved along the directions of the major and
minor axes of the relative orbit, and
$$
\br = a(\cos(g+l) - {3\over2}e\cos g + {1\over 2}e \cos(2l+g), \sin(g+l)
- {3\over2}e\sin g + {1\over2}e\sin(2l+g)),\eqno\stepeq
$$
relative to the same frame as that in which $g$ is measured.  The effect
of the transformation $g\rightarrow g+\pi$ is obvious.

Because $g$ is ill-defined when $e$ becomes very small, it is better to
use the Poincar\'e variables $L^\prime$, $l^\prime$, $\xi$ and $\eta$,
defined by $L^\prime = L,~l^\prime = l+g,$
$$~\xi = \sqrt{2G^\prime}\cos g\eqno\stepeq
$$ 
and 
$$
\eta = -\sqrt{2G^\prime}\sin g\eqno\stepeq
$$ 
where
$$
G^\prime = L - G \simeq (1/2)e^2L,\eqno\stepeq
$$ 
by equation~(B2).  These are still
canonical, $l^\prime$ and $\eta$ being conjugate to $L^\prime$ and
$\xi$, respectively.  In these variables equation~(B5) becomes 
$$ 
\br = a(\,\cos
l^\prime - {3\over2}{\xi\over\sqrt{L^\prime}}
+{1\over2\sqrt{L^\prime}}(\xi\cos2l^\prime - \eta\sin2l^\prime)\, , \,\sin
l^\prime + {3\over2}{\eta\over\sqrt{L^\prime}} 
+{1\over2\sqrt{L^\prime}}(\xi\sin2l^\prime + \eta\cos2l^\prime)\,)\eqno\stepeq
$$
approximately,
and the transformation $l^\prime\rightarrow l^\prime+\pi$, 
$\xi\rightarrow-\xi$, $\eta\rightarrow-\eta$ sends $\br\rightarrow-\br$.

In general, $\br$ is a $2\pi$-periodic function of $l^\prime$, and can be
expanded as a Fourier series $\br = \sum_n
\br_n(L^\prime,\xi,\eta)\exp(inl^\prime)$, where the coefficients
$\br_n$ are actually power series in $\xi$ and $\eta$.  This is
illustrated at lowest non-trivial order by equation~(B9).  Similarly, we may
expand the perturbing function as a series
$\Re = \sum_{-\infty}^\infty
a_n(L^\prime,\xi,\eta,\bR)\exp(inl^\prime),$ where again the
coefficients are power series in the variables $\xi$ and $\eta$.  Now
the invariance of the Hamiltonian under the transformation $\br\to-\br$
(for equal masses)
implies that this series is invariant under the transformation
$l^\prime\to l^\prime+\pi$, $\xi\to-\xi$, $\eta\to-\eta$.  If the
coefficient $a_n$ contains a term $\xi^\alpha\eta^\beta$ it follows that
$\alpha+\beta$ is even if $n$ is even, and  $\alpha+\beta$ is odd if $n$ is
odd.

The aim of Hamiltonian perturbation theory is to remove those terms
which depend on quickly-varying angles.  At lowest order, this is
equivalent to averaging over the fast motions.  Since $l$ or $l^\prime$ 
is the one
rapidly changing quantity, this leaves only those terms in the Fourier
series which have $n = 0$.  These terms have as coefficients power
series in $\xi$ and $\eta$ which have even ``order" (i.e. the value of
$\alpha+\beta$).  When these are differentiated with respect to these
variables (to give Hamilton's equations for their evolution)  the
resulting expression has terms of only {\sl odd} order in these
variables.  It follows, therefore, that one solution is $\xi = \eta =
0$, and since this solution yields the correct initial condition for a
circular binary, it follows that the induced eccentricity vanishes (in
first order perturbation theory).

All we have to do now is to show that terms of the wrong parity will not
be introduced in the perturbation procedure.  In the standard procedure,
which is described in many texts (e.g. Born 1960; Goldstein 1980;
Lichtenberg \& Liebermann 1983), the
terms with $n\ne0$ are removed by a near-identity canonical
transformation.  For example, a term $a_n\exp(inl^\prime)$ is removed by
a transformation to new variables (denoted here by stars) in which
$$
l^{\prime\star} = l^\prime + \displaystyle{{\partial S_1\over\partial
L^{\prime\star}}e^{inl^\prime}},~~~~~~~~
\xi= \xi^\star  + \displaystyle{{\partial
S_1\over\partial\eta}e^{inl^\prime}},\eqno\stepeq
$$ 
etc., where
$$
S_1(\eta,\bR,L^{\prime\star},\xi^\star,\bP^\star) = -
\displaystyle{{a_n(\eta,\bR,L^{\prime\star},\xi^\star,\bP^\star)\over
in\displaystyle{{\partial H_b(L^{\prime\star})\over\partial
L^{\prime\star}}}}}\eqno\stepeq
$$ 
(cf.\ Lichtenberg \& Lieberman 1983, eq.~[2.2.17]).  

Now it is clear from inspection of these equations 
(and the corresponding equations for the transformation of the other
variables), 
along with our result on the parity of the coefficient $a_n$,
that the transformation $l^\prime\to l^\prime+\pi$,
$\xi\to-\xi$, $\eta\to-\eta$ corresponds, in the new variables, to the
transformation $l^{\prime\star}
\to l^{\prime\star}+\pi$,
$\xi^\star\to-\xi^\star$, $\eta^\star\to-\eta^\star$.  Since the
Hamiltonian was invariant under the transformation in the old variables,
it will also be invariant under the transformation in the new ones.  It
follows that, when the Hamiltonian is expressed in terms of the new
variables, the new coefficients $a_n$ will have the same parity property
as the old ones.  It follows that the induced eccentricity will again
vanish for an initially circular binary.  By induction, this conclusion
follows for perturbation theory at all orders.

\end